%% file: main.tex
\documentclass[11pt,a4paper]{scrartcl}

\usepackage[utf8]{inputenc}
\usepackage[english]{babel}           
\usepackage{amsfonts, amsmath, amsthm, amssymb, dsfont}

\usepackage{newtxmath}

\usepackage{graphicx}
\usepackage{epstopdf}
\usepackage{color}
\usepackage{hyperref}

\usepackage{lineno}

\usepackage{natbib}

\usepackage{tikz} % diagrams
\usetikzlibrary{arrows.meta,arrows}
\usetikzlibrary{positioning}
\newcommand{\AxisRotator}[1][rotate=0]{%
    \tikz [x=0.25cm,y=0.60cm,line width=.2ex,-Latex,#1] \draw (0,0) arc (-150:150:1 and 1);%
}

\usepackage{epstopdf}
\AppendGraphicsExtensions{.tiff}

\DeclareMathAlphabet\mathsfbi{T1}{phv}{b}{it}

\makeatletter
\DeclareOldFontCommand{\rm}{\normalfont\rmfamily}{\mathrm}
\DeclareOldFontCommand{\sf}{\normalfont\sffamily}{\mathsf}
\DeclareOldFontCommand{\tt}{\normalfont\ttfamily}{\mathtt}
\DeclareOldFontCommand{\bf}{\normalfont\bfseries}{\mathbf}
\DeclareOldFontCommand{\it}{\normalfont\itshape}{\mathit}
\DeclareOldFontCommand{\sl}{\normalfont\slshape}{\@nomath\sl}
\DeclareOldFontCommand{\sc}{\normalfont\scshape}{\@nomath\sc}
\makeatother

\title{An unstable mode of the stratified atmosphere under the non-traditional Coriolis acceleration}
\author{Ray Chew$^{\text{(a)}}$, Mark Schlutow$^{\text{(b)}}$\footnote{Corresponding author: Mark Schlutow, mark.schlutow@bgc-jena.mpg.de}~\footnote{M. Schlutow and R. Chew are co-first authors of this paper.}~, Rupert Klein$^{\text{(c)}}$}
\date{}

\begin{document}

	\maketitle	
\begin{center}
\vspace{-1.5cm}
{
$^{\text{(a)}}$\,Institut für Atmosphäre und Umwelt \\
Goethe Universität Frankfurt am Main \\
Altenhöferallee 1, 60438 Frankfurt am Main, Germany
}
{
\vskip 0.5cm
$^{\text{(b)}}$\,Department Biogeochemical Signals \\
Max Planck Institute for Biogeochemistry \\
Hans-Kn\"oll-Str. 10, 07745 Jena, Germany
}
{
\vskip 0.5cm
$^{\text{(c)}}$\,FB Mathematik \& Informatik \\
Freie Universit\"at Berlin \\
Arnimallee 6, 14195 Berlin, Germany
}
\end{center}	

	\newpage

	\begin{abstract}
The traditional approximation neglects the cosine components of the Coriolis acceleration, and this approximation has been widely used in the study of geophysical phenomena. However, the justification of the traditional approximation is questionable under a few circumstances. In particular, dynamics with substantial vertical velocities or geophysical phenomena in the tropics have non-negligible cosine Coriolis terms. Such cases warrant investigations with the non-traditional setting, i.e., the full Coriolis acceleration. In this manuscript, we study the effect of the non-traditional setting on an isothermal, hydrostatic and compressible atmosphere assuming a meridionally homogeneous flow. Employing linear stability analysis, we show that, given appropriate boundary conditions, i.e., a bottom boundary condition that allows for a vertical energy flux and non-reflecting boundary at the top, the atmosphere at rest becomes prone to a novel unstable mode. The validity of assuming a meridionally homogeneous flow is investigated via scale analysis. Numerical experiments were conducted, and Rayleigh damping was used as a numerical approximation for the non-reflecting top boundary. Our three main results are as follows. 1) Experiments involving the full Coriolis terms exhibit an exponentially growing instability, yet experiments subjected to the traditional approximation remain stable. 2) The experimental instability growth rate is close to the theoretical value. 3) A perturbed version of the unstable mode arises even under sub-optimal bottom boundary conditions. Finally, we conclude our study by discussing the limitations, implications, and remaining open questions. Specifically, the influence on numerical deep-atmosphere models and possible physical interpretations of the unstable mode are discussed.
	\end{abstract}

	\hrule
	
	\hspace{8mm}

	\vspace{1cm}

	%\hspace{3mm}
	
\section{Introduction}
The Coriolis force arises from the Earth's rotation. Named after \citet{coriolis1835memoire}, the effect of the Coriolis force is the deflection of a particle moving along the surface of the globe. Such an effect is perpendicular to the axis of the globe's rotation and the direction of the particle's velocity. Atmospheric and oceanic phenomena are subjected to the effects of the Coriolis force. Yet, as the atmosphere and ocean are shallow relative to the Earth's radius, the dynamics we are interested in are largely horizontal.

Based on the argument of the difference in the horizontal and vertical scales of the dynamics, \citet{laplace1835oeuvres} concluded that in studying the Coriolis effects, we may neglect the vertical acceleration arising from the Coriolis force, keeping only the horizontal. Specifically, the full Coriolis acceleration results in a tilted outward-pointing vector along the surface of the globe. This vector may be decomposed into sine and cosine components. The former are perpendicular to the surface of the globe and interact with only horizontal motion, while the latter are parallel and associated with vertical motion. Appendix~\ref{apx:coriolis} provides an elaboration on the full Coriolis acceleration. Following the arguments by Laplace, one may drop the cosine terms.

The approximation introduced by Laplace has since been widely used. This led \citet{eckart1960hydrodynamics} to coin the term ``traditional approximation'' when describing this approximation. The traditional approximation is also introduced in other standard textbooks on atmospheric and oceanic phenomena, see e.g., \citet{pedlosky2013geophysical,vallis2017atmospheric,achatz2022atmospheric}. However, the validity of the traditional approximation may be questioned under certain circumstances. For instance, at the equator where the contributions from the cosine terms are the largest, or for phenomena involving substantial vertical motion, wherein the vertical Coriolis acceleration should not be ignored.

The limitations of the traditional approximation are gaining increasing attention, and the review by \citet{gerkema_geophysical_2008} may be a good starting point for a study beyond the traditional approximation. A few other relevant references to the literature are mentioned below. \citet{white_dynamically_1995} proposed a set of dynamically consistent quasi-hydrostatic equations with full Coriolis support. They also demonstrated via scale analysis that, in the tropics, the cosine Coriolis terms may be as large as 10\% of the important terms in the equations, and thus they are non-negligible. The condition for linear stability has been derived by \citet{fruman_symmetric_2008} for a zonally symmetric, compressible, and non-traditional setup. They noticed that an instability arises for a particular angular momentum profile near the equator. \citet{fruman_equatorially_2009} computed the analytical solution for meridionally-confined zonally-propagating waves for the linearised hydrostatic Boussinesq equations including the full Coriolis terms. Studies involving non-traditional effects were also conducted by \citet{de1994flows,maas2007equatorial,igel2020nontraditional,rodal_waves_2021-1,perez_unidirectional_2022}.

Challenges to the assumptions made in the traditional approximation are not limited to the analysis of physical phenomena. In the example of a numerical model, \citet{borchert_upper-atmosphere_2019} introduced an upper-atmosphere extension to the ICON model \citep{zangl_icon_2015} that includes the full Coriolis term. Another deep-atmosphere numerical model with full Coriolis support has been published by \citet{smolarkiewicz_finite-volume_2016}. From hereon, we use the term ``non-traditional setting'' to refer to a setting where the full effect of the Coriolis force is considered.

In this manuscript, we study the stability of an isothermal hydrostatic atmosphere under the non-traditional setting and assuming a meridionally homogeneous flow. Via linear stability analysis of the compressible inviscid Euler equations, we show that, given the non-reflective boundary condition at the top and a forcing boundary at the bottom, a small perturbation of the isothermal hydrostatic atmosphere at rest leads to an exponentially-growing instability. We identify the structure of this unstable mode and quantify its theoretical growth rate. 
In contrast to this choice of boundary conditions, it can be shown by an argument from functional analysis in combination with energy conservation that the atmosphere is unconditionally linearly stable if a free-slip boundary condition is assumed at the flat top and bottom boundaries. Section~\ref{sec:theory} contains these theoretical developments. The assumption of meridionally homogeneous flow is also discussed in detail there.

For the case of a non-reflecting upper boundary and a bottom forcing that allows for a vertical flux of energy into the domain, the unstable mode is investigated through numerical experiments. Section~\ref{sec:num_model} and Appendix~\ref{apx:num_model} introduce the numerical model by \citet{bk2019}, and {Appendix~\ref{apx:extension}} extends the numerical model to support the non-traditional setting. Appendix~\ref{apx:lamb_bal} details the necessary numerical extension to ensure a well-balanced Lamb wave solution, and Appendix~\ref{apx:energy} establishes the energy-conserving properties of the stable Lamb wave solution by the numerical model. The non-reflective top boundary is realised numerically via a Rayleigh damping layer.

Experiments made in Section~\ref{sec:num_experiments} demonstrate that, in the presence of the full Coriolis acceleration, the isothermal, hydrostatic atmosphere at rest becomes unstable with an experimental instability growth rate that is close to the theoretically predicted value. The instability growth is also observed in an experiment where the bottom forcing applied deviates from the one that was used to closely represent the analytical theory within the numerical simulation framework.
Results with this sub-optimal forcing suggest that related unstable modes exist for a variety of bottom boundary conditions.
In nature, the atmospheric boundary layer flow will provide effective bottom boundary conditions, for instance, and in conjunction with non-homogeneous surface conditions, these may induce an effective bottom forcing akin to the sub-optimal forcing employed here.

The existence of this unstable mode due to the full Coriolis effect has two main implications. First, this unstable mode may provide a theoretical understanding of hitherto unexplained physical phenomena. Second, the presence of the unstable mode is relevant to numerical weather and climate prediction models that solve a full representation of the Coriolis force, in particular numerical models of the deep-atmosphere. Along with a summary of the results of this manuscript, Section~\ref{sec:conclusions} provides further elaborations and discussions of these implications.

\section{Theory}
\label{sec:theory}
\subsection{The governing equations}
The basis for our investigation are the compressible Euler equations in a rotating coordinate system on the sphere, reproduced below,
\begin{subequations}
	\label{eq:govern}
	\begin{align}
		&\frac{D\pmb{v}}{Dt}+c_p\,\theta\nabla\pi+g\,\pmb{e}_z
		+2\,\pmb{\Omega}\times\pmb{v}=0,\\
		&\frac{D\theta}{Dt}=0,\\
		&\frac{D\pi}{Dt}+\frac{R}{c_v}\,\pi\nabla\cdot\pmb{v}=0,
		\label{eq:govern_c}
	\end{align}
\end{subequations}
where the velocity vector $\pmb{v}=u\,\pmb{e}_x+v\,\pmb{e}_y+w\,\pmb{e}_z$ is composed of the zonal, meridional and vertical wind. The variable $R=c_p-c_v$ denotes the specific gas constant for dry air, where $c_p$ and $c_v$ are the heat capacities at constant pressure and constant volume, respectively.
Furthermore, $\pmb{e}_j,~j\in\{x,y,z\}$ represents the unit vectors, and $g$ is the gravitational acceleration. 
The Earth's angular velocity vector is $\pmb{\Omega} = \Omega_x \pmb{e}_x + \Omega_y \pmb{e}_y + \Omega_z \pmb{e}_z$.
The Exner pressure $\pi$ and potential temperature $\theta$ are defined by the canonical thermodynamic quantities, temperature $T$ and pressure $p$. These quantities are related by the equations of state,
\begin{subequations}
\label{eq:thermo}
	\begin{align}
		\pi&=(p/p_0)^\kappa,\\
		\theta&=T/\pi,
	\end{align}
\end{subequations}
with $p_0$ a reference pressure at $z=z_0$ and $\kappa=R/c_p=2/7$ for two-atomic gases. By means of the ideal gas law, the thermodynamical variables are linked to the density $\rho$ via
\begin{align}
	\label{eq:ideal}
	\rho=\frac{p_0}{R}\frac{\pi^{(1-\kappa)/\kappa}}{\theta}.
\end{align}
The material derivative is given as
\begin{align}
	\frac{D}{Dt}=\frac{\partial}{\partial t}+\pmb{v}\cdot\nabla
\end{align}
where $\nabla = \pmb{e}_x\,\partial / \partial x + \pmb{e}_y\,\partial / \partial y + \pmb{e}_z\,\partial / \partial z$ denotes the nabla operator. Finally, the vector $\pmb{\Omega}$ represents Earth's angular velocity in the direction of the Earth's rotational axis.
The governing equations introduced in \eqref{eq:govern} are based on mass, momentum, and energy conservation.

\subsection{The non-traditional setting}
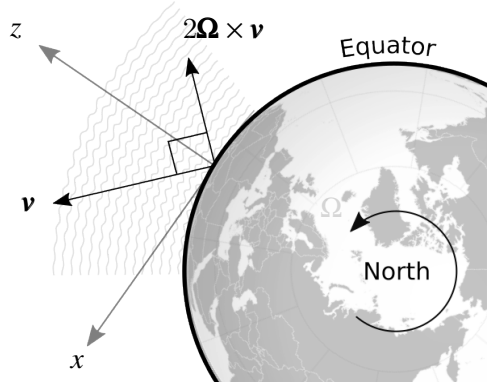
\begin{figure}
	\centerline{\input{SketchOfSetup.tex}}
	\caption{Sketch of the Coriolis acceleration at the equator.}
	\label{fig:sketch}
\end{figure}

The effect of the non-traditional setting is most obvious close to the equator, and so we assume $f$-$F$-planes at the equator \citep{thuburn_normal_2002}, i.e., $f=0$ and $F=2\Omega$ with $\Omega=|\pmb{\Omega}|$. More details on the $f$-$F$-planes notation may be found in Appendix~\ref{apx:coriolis}. The $\beta$-plane effect is additionally discussed by means of scale analysis in Section~\ref{subsec:beta_influence}. From hereon, we consider only the flow in the $x$-$z$-plane, i.e., we restrict the analysis to meridionally homogeneous fields. The horizontal domain of interest is $x\in[0,x_{\rm MAX}]$ with $z\in[z_0,z_\mathrm{TOA}]$ for the vertical extent and $z_0$ the reference height. TOA stands for a \textit{fictitious} demarcation representing the `top of the atmosphere', and a clearer elaboration on the boundary conditions will be provided in Subsection~\ref{subsec:linear_analysis}. This problem setup is illustrated in Figure~\ref{fig:sketch}.

Writing out the compact form of \eqref{eq:govern} explicitly yields the following,
\begin{subequations}
\label{eq:simplified_govern}
    \begin{align}
        &\frac{\partial u}{\partial t}
        +u\frac{\partial u}{\partial x}
        +w\frac{\partial u}{\partial z}
        +c_p\theta\frac{\partial\pi}{\partial x}
        +Fw=0,\\
        \label{eq:govern_w}
        &\frac{\partial w}{\partial t}
        +u\frac{\partial w}{\partial x}
        +w\frac{\partial w}{\partial z}
        +c_p\theta\frac{\partial\pi}{\partial z}
        -Fu+g=0,\\
        &\frac{\partial \theta}{\partial t}
        +u\frac{\partial \theta}{\partial x}
        +w\frac{\partial \theta}{\partial z}=0,\\
        &\frac{\partial \pi}{\partial t}
        +u\frac{\partial \pi}{\partial x}
        +w\frac{\partial \pi}{\partial z}
        +\frac{R}{c_v}\pi\left(
        \frac{\partial u}{\partial x}
        +\frac{\partial w}{\partial z}
        \right)=0.
    \end{align}
\end{subequations}

\subsection{The hydrostatic background atmosphere}
A stationary solution to the governing equations \eqref{eq:simplified_govern} is the hydrostatic atmosphere at rest, i.e.,
\begin{subequations}
\begin{align}
	u(x,z,t)&=0,\\
	w(x,z,t)&=0,\\
	\theta(x,z,t)&=\bar\theta(z),\\
	\pi(x,z,t)&=\bar\pi(z).
\end{align}
\label{eqn:hydrostatic_approximation}%
\end{subequations}%
Substituting \eqref{eqn:hydrostatic_approximation} into \eqref{eq:govern}, all governing equations but the vertical velocity equation \eqref{eq:govern_w} vanish. The remaining equation reduces to the hydrostatic equation,
\begin{align}
    \label{eq:hydr_stat}
	c_p\,\bar\theta\frac{\mathrm{d}\bar\pi}{\mathrm{d}z}=-g.
\end{align}
Given a temperature profile $\bar{T}=\bar{T}(z)$, we can integrate \eqref{eq:hydr_stat} using the definitions in \eqref{eq:thermo} to obtain the hydrostatic background variables, and they may be written as follows,
\begin{align}
    \bar{\pi}(z)&=\exp\left[-\int_0^{z}\frac{g}{c_p}\frac{1}{\bar{T}(\zeta)}\mathrm{d}\zeta\right],\\
    \bar{\theta}(z)&=T_0\exp\left[\int_0^{z}\frac{1}{\bar{T}(\zeta)}\left(\frac{g}{c_p}
	+\frac{\mathrm{d}\bar{T}}{\mathrm{d}\zeta}\right)\mathrm{d}\zeta\right],\\
    \bar\rho(z)&=\frac{p_0}{RT_0}\exp\left[-\int_0^{z}\frac{1}{\bar{T}(\zeta)}\left(\frac{g}{R}
	+\frac{\mathrm{d}\bar{T}}{\mathrm{d}\zeta}\right)\mathrm{d}\zeta\right].
\end{align}

Subsequently, we consider perturbations from the basic, hydrostatic state by the following ansatz,
\begin{subequations}
\begin{align}
	u(x,z,t)&=u'(x,z,t),\\
	w(x,z,t)&=w'(x,z,t),\\
	\theta(x,z,t)&=\bar\theta(z)+\theta'(x,z,t),\\
	\pi(x,z,t)&=\bar\pi(z)+\pi'(x,z,t).
	\label{eqn:pi_perturb}
\end{align}
\label{eqn:perturbations}%
\end{subequations}%
Assuming that the perturbations are infinitesimally small, the governing equations may be linearised around the basic hydrostatic state.
Before stating the linearised system, we introduce a transformation of the primed variables concatenated into a vector $\pmb{\chi}=(\chi_u,\,\chi_w,\,\chi_\theta,\,\chi_\pi)^\mathrm{T}$ 
that simplifies the upcoming derivations \citep[cf.][Eq.~2.16]{achatz_gravity_2010},%
\begin{align}
    \pmb{\chi}=\bar{\rho}^{1/2}\left(u',\,w',\,
    \frac{g}{N}\,\frac{{\theta^\prime}}{\bar{\theta}},\,
    \frac{c_p}{C}\,\bar{\theta}\,\pi'\right)^\mathrm{T}.
\label{eqn:hat_eqns}
\end{align}
Here, the Brunt-Väisälä frequency and the speed of sound are given by
\begin{align}
	N(z)&=\sqrt{\frac{g}{\bar{\theta}(z)}\frac{\mathrm{d}\bar{\theta}}{\mathrm{d}z}}
	=\sqrt{\frac{g}{\bar{T}(z)}\left(\frac{g}{c_p}
	\label{eqn:bv_freq}
	+\frac{\mathrm{d}\bar{T}}{\mathrm{d}z}\right)},\\
	C(z)&=\sqrt{\gamma R\bar{T}(z)},\quad \gamma=\frac{c_p}{c_v},
\end{align}
respectively. Inserting \eqref{eqn:hat_eqns} into \eqref{eqn:perturbations} and the result into \eqref{eq:simplified_govern}, we may write the linearised system in terms of the transformed perturbation vector components as
\begin{subequations}
	\label{eq:lin}
	\begin{align}
		&\frac{\partial\chi_u}{\partial t}+\frac{\partial}{\partial x}\bigl(C\chi_\pi\bigr)+F\chi_w=0,\\
		&\frac{\partial\chi_w}{\partial t}+\left(\frac{\partial}{\partial z}+\Gamma\right)\bigl(C\chi_\pi\bigr)-N\chi_\theta-F\chi_u=0,
		\\
		&\frac{\partial\chi_\theta}{\partial t}+N\chi_w=0,\\
		&\frac{\partial\chi_\pi}{\partial t}+C\left[\frac{\partial\chi_u}{\partial x}+\left(\frac{\partial}{\partial z}-\Gamma\right)\chi_w\right]=0,
	\end{align}
\end{subequations}
where we introduced the quantity $\Gamma$, which has dimension of inverse height, as
\begin{align}
	\Gamma(z)&=-\frac{1}{2\bar{\rho}(z)}\frac{\mathrm{d}\bar{\rho}}{\mathrm{d}z}
	-\frac{1}{\bar{\theta}(z)}\frac{\mathrm{d}\bar{\theta}}{\mathrm{d}z} \nonumber \\
	&=\frac{1}{2\bar{T}(z)}\left[\left(\frac{c_v}{R}-1\right)\frac{g}{c_p}
	-\frac{\mathrm{d}\bar{T}}{\mathrm{d}z}\right]
	\label{eqn:Gamma}
\end{align}
in accordance with \citet{durran_improving_1989}. Notice that by the choice of the variable transformation,
\begin{align}
    e=\frac{1}{2} \left( \chi_u^2+\chi_w^2+\chi_\theta^2+\chi_\pi^2 \right)
\end{align}
is the total energy density of the perturbation which evolves in time as
\begin{align}
    \label{eq:energy_density_conservation}
    \frac{\partial e}{\partial t}+\frac{\partial}{\partial x}\bigl(C\chi_u\chi_\pi\bigr)
    +\frac{\partial}{\partial z}\bigl(C\chi_w\chi_\pi\bigr)=0.
\end{align}
Hence, Gauss' theorem ensures that the total energy is locally conserved. The quantity $\chi_u^2+\chi_w^2$ represents the kinetic energy density of the perturbation, and the quantity $\chi_\theta^2 + \chi_\pi^2$ represents the sum of the potential and internal energy density.
The perspective of energy conservation lends an additional physical interpretation to the newly introduced variable $\Gamma$. 
Considering \eqref{eq:lin}, we observe that the quantity $C\Gamma$ represents an energy exchange rate for the conversion between kinetic energy due to vertical motion and internal energy which is associated with the compressibility of the gas.

\subsection{Linear stability analysis}
\label{subsec:linear_analysis}
Proceeding with our theoretical developments, notice that the solution of the linearised equations \eqref{eq:lin} is of the form
\begin{align}
    \pmb{\chi}(x,z,t)=\pmb{\psi}(x,z)\,\exp(-i\lambda t),\quad \lambda\in\mathbb{C}.
    \label{eqn:soln_ansatz}
\end{align}
This solution ansatz yields an eigenvalue problem with eigenvalues $i\lambda$, i.e., 
\begin{align}
    (\mathbf{T}-i\lambda)\,\pmb{\psi}=0,
    \label{eqn:eigenvalue_problem}
\end{align}
with the differential operator
\begin{align}
    \mathbf{T}=
    \begin{pmatrix}
    0& F& 0& \partial/\partial x(C~\cdot~)\\
    -F& 0& -N& \partial/\partial z(C~\cdot~)+\Gamma C\\
    0& N& 0& 0\\
    C\partial/\partial x& C\partial/\partial z-C\Gamma& 0& 0
    \end{pmatrix}.
\end{align}
Note that as $\pmb{\psi}$ is complex valued, the solution of \eqref{eqn:eigenvalue_problem} is a linear combination of the eigenfunctions. This ensures that the solutions in \eqref{eqn:soln_ansatz} are eventually real fields.

Now, let us define an inner product as
\begin{align}
    \langle\pmb{\psi}|\pmb{\phi}\rangle
    =\int_{z_0}^{z_\mathrm{TOA}}\int_0^{x_\mathrm{MAX}}
    \pmb{\psi}^\mathrm{H}\pmb{\phi}
    ~\mathrm{d}x\mathrm{d}z
\end{align}
in terms of some well-behaved vectors of prognostic variables
$\pmb{\psi}=(\psi_u,\,\psi_w,\,\psi_\theta,\,\psi_\pi)^\mathrm{T}$ 
and $\pmb{\phi}=(\phi_u,\,\phi_w,\,\phi_\theta,\,\phi_\pi)^\mathrm{T}$
that satisfy the yet to be defined boundary conditions.
Furthermore, $\mathrm{H}$ denotes the Hermitian transpose, i.e the conjugate transpose. This inner product induces a norm that is given as
\begin{align}
    \|\pmb{\psi}\|^2
    =\langle\pmb{\psi}|\pmb{\psi}\rangle
    =\int_{z_0}^{z_\mathrm{TOA}}\int_0^{x_\mathrm{MAX}}
    \pmb{\psi}^\mathrm{H}\pmb{\psi}
    ~\mathrm{d}x\mathrm{d}z
    =\int_{z_0}^{z_\mathrm{TOA}}\int_0^{x_\mathrm{MAX}} 2\,e
    ~\mathrm{d}x\mathrm{d}z = 2E,
    \label{eqn:2-norm}
\end{align}
which corresponds to twice the global energy E, and therefore this choice of the inner product is motivated by a physical argument.

At this point, an open question that we still have to consider is the choice of realistic boundary conditions. To answer this question, let us consider the properties of the operator $\mathbf{T}$. If we assume for simplicity periodic boundary conditions in the $x$-direction, such that $\pmb{\psi}(0,z)=\pmb{\psi}({x_\mathrm{MAX}},z)$ (or vanishing fields at $x\rightarrow\pm\infty$), we then observe that
\begin{align}
    \langle\mathbf{T}\boldsymbol{\psi}|\boldsymbol{\phi}\rangle
    =-\langle\boldsymbol{\psi}|\mathbf{T}\boldsymbol{\phi}\rangle
    +\int_0^{x_\mathrm{MAX}}\Bigl[ 
    C\psi_\pi^\ast\phi_w+C\psi_w^\ast\phi_\pi
    \Bigr]_{z_0}^{z_\mathrm{TOA}}
    \,\mathrm{d}x
\end{align}
for all $\pmb{\psi}$ and $\pmb{\phi}$. Here, $\ast$ represents the complex conjugate. This observation implies that the operator $\mathbf{T}$ is skew-Hermitian if the following condition holds,
\begin{align}
    \label{eq:leftover}
    \Bigl[C\psi_\pi^\ast\phi_w+C\psi_w^\ast\phi_\pi
    \Bigr]_{z_0}^{z_\mathrm{TOA}}=0.
\end{align}
Skew-Hermitian operators exhibit purely imaginary eigenvalues $i\lambda$ (or rather an imaginary spectrum to be precise), so $\lambda\in\mathbb{R}$. Therefore, the perturbations in \eqref{eq:lin} are bounded in time and ultimately stable if the condition in \eqref{eq:leftover} holds.

Whether or not condition \eqref{eq:leftover} holds now depends on the choice of the top and bottom boundaries. If we were to assume a free-slip (no-normal-flow) boundary, i.e., $w(x,0,t)=z_0$, for the bottom with no orography, then this boundary condition implies that
\begin{align}
    \psi_w,\,\phi_w=0\quad\text{at }z=z_0.
\end{align}
If we were to also restrict the fields to vanish at the upper boundary, then the condition \eqref{eq:leftover} holds, and all perturbations would be stable. A similar result generalised to the equatorial $\beta$-plane was also obtained by \citet{fruman_symmetric_2008} who considered zonally symmetric flows. 
In fact, all trivial boundary conditions would result in stable perturbations. We can illuminate this statement by integrating \eqref{eq:energy_density_conservation}
over the spatial domain, which leaves us with the rate of change of the global energy $E$ given in terms of the vertical fluxes of total energy 
\begin{align}
    \label{eq:roc_glob_energy}
    \frac{\mathrm{d}E}{\mathrm{d}t}=\int_0^{x_\mathrm{MAX}}\bigl.C\chi_w\chi_\pi\bigr|_{z=z_0}\,\mathrm{d}x-\int_0^{x_\mathrm{MAX}}\bigl.C\chi_w\chi_\pi\bigr|_{z=z_\mathrm{TOA}}\,\mathrm{d}x
\end{align}
 at $z_0$ and the top of the atmosphere, respectively.
Therefore, for an instability to grow in energy, there must be an energy flux through the boundaries, but for trivial boundary conditions, the flux would vanish. As energy cannot come from space, the only reasonable energy source must be at $z_0$.

\subsection{The isothermal background atmosphere}
Let us now simplify our model even further by assuming an isothermal background state, $\partial\bar{T}/\partial z=0$. As a consequence, all coefficients in $\mathbf{T}$ become constant and, taking into account the horizontal boundary conditions, the perturbations must now be of the form
\begin{align}
    \label{eq:pert_ansatz}
    \pmb{\psi}(x,z)=\tilde{\pmb{\psi}}\,\exp(ikx+\mu z),\quad k\in\mathbb{R}, \mu\in\mathbb{C},
\end{align}
where $\tilde{\pmb{\psi}}$ is a vector with constant components.
The operator $\mathbf{T}$ may be written as a matrix, that is,
\begin{align}
\tilde{\mathbf{T}}=
	\begin{pmatrix}
		0 & F & 0 & iCk\\
		-F & 0 & -N & C(\mu+\Gamma)\\
		0 & N & 0 & 0\\
		iCk & C(\mu-\Gamma) & 0 & 0
	\end{pmatrix}.
	\label{eqn:T_matrix}
\end{align}%
Dividing \eqref{eqn:T_matrix} by $N$, the system may be non-dimensionalised by defining new dimensionless variables that are given as follows,
\begin{align}
	\label{eq:nondim}
	K=\frac{C}{N}\,k,\quad M=\frac{C}{N}\,\mu,\quad
	\Lambda=\frac{1}{N}\,\lambda,\quad
	\varepsilon=\frac{F}{N}.
\end{align}

Writing the eigenvalue problem of \eqref{eqn:eigenvalue_problem} in terms of the dimensionless variables and incorporating the horizontal periodic boundary conditions leads to
\begin{align}
\bigl(\mathbf{S}-i\Lambda\bigr)\,\tilde{\pmb{\psi}}=0,\quad
\mathbf{S}(K,M;\varepsilon)=
	\begin{pmatrix}
		0 & \varepsilon & 0 & iK\\
		-\varepsilon & 0 & -1 & M+G\\
		0 & 1 & 0 & 0\\
		iK & M-G & 0 & 0
	\end{pmatrix},
\end{align}
where we have
\begin{align}
    G&=\frac{C\Gamma}{N}=\frac{1-\gamma/2}{\sqrt{\gamma-1}}=\sqrt{\frac{9}{40}}.
\end{align}
The characteristic polynomial of $\mathbf{S}$ is obtained, and it is
\begin{align}
   	\label{eq:charpol}
	\mathcal{P}_{\mathbf{S}(K,M;\,\varepsilon)}(i\Lambda)=\Lambda^4
	-\left(1+\varepsilon^2+G^2+K^2-M^2\right)\,\Lambda^2
	+2\varepsilon \,G K\,\Lambda
	+K^2.
\end{align}
As this polynomial is a depressed quartic function in $\Lambda$, its roots may be expressed by radicals. However, these expressions would be tedious and provide no further elucidation. Instead, we note that $\varepsilon=\mathit{O}\left(10^{-2}\right)$ is typically a very small number, and so it may be sensible to employ perturbation theory for the subsequent developments. If we assume that $M,\,K=\mathit{O}(1)$ as $\varepsilon\rightarrow 0$, then we have, to leading order, the following roots for the characteristic polynomial:
\begin{align}
    \label{eq:dispersion0}
	{\Lambda^{(0)}}^2&=\frac{1}{2}L^2\left(1\pm\sqrt{1-4\frac{K^2}{L^4}}\right),\\
	L^2&=K^2-M^2+1+G^2=K^2-M^2+\frac{49}{40},
\end{align}
which is essentially the dispersion relation for acoustic-gravity waves.

\subsection{The leading-order eigenvector}
Computing the leading-order eigenvector of $\mathbf{S}(K,M;0)$ gives us
\begin{align}
    \tilde{\pmb{\psi}}^{(0)}=
    \left(1,\, \frac{i}{K}\frac{{\Lambda^{(0)}}^2-K^2}{M-G},\,
    \frac{1}{{\Lambda^{(0)}} K}\frac{{\Lambda^{(0)}}^2-K^2}{M-G},\,
    \frac{{\Lambda^{(0)}}}{K}\right)^\mathrm{T}.
    \label{eqn:leading_order_eigenvector}
\end{align}
A particularly interesting situation arises when we assume a free-slip boundary condition at the ground to the leading order, such that $\psi_w^{(0)}=0$ at $z=z_0$,
because the leading-order total energy flux then also vanishes according to \eqref{eq:roc_glob_energy}.
If we were able to show that an instability occurs in the next-order correction, then the energy flux would be, at most, next-to-leading order.
Or, in other words, a minuscule energy flux through the bottom boundary that would otherwise be ignored might lead to an exponentially growing perturbation.
We consider this situation worthwhile to investigate in more detail.
The leading order eigenvector \eqref{eqn:leading_order_eigenvector} fulfils the free-slip boundary condition non-trivially if 
\begin{align}
    {\Lambda^{(0)}}^2=K^2.
    \label{eqn:Lambda_K_rs}
\end{align}
Inserting \eqref{eqn:Lambda_K_rs} into \eqref{eq:charpol}, we can solve the leading-order characteristic polynomial to obtain 
\begin{align}
    M=-G,
\end{align}
which ensures a physical solution that decays as $z\rightarrow\infty$.
Hence, the leading-order eigenvector becomes
\begin{align}
    \label{eq:lead_ord_eig_vec}
    \tilde{\pmb{\psi}}^{(0)}=
    \bigl(1,\, 0,\, 0,\, \pm 1\bigr)^\mathrm{T},
\end{align}
and this solution describes a Lamb wave \citep{vallis2017atmospheric}. Since the characteristic polynomial is a quartic, we expect to find four roots. Yet for the free-slip lower boundary condition we imposed, only two roots were obtained in \eqref{eqn:Lambda_K_rs}. We note that the leading-order polynomial has two more roots for $M=-G$ that do not satisfy the leading-order boundary conditions for all $K$, and these are
\begin{align}
    {\Lambda^{(0)}}^2=1.
\end{align}
These roots would describe Brunt waves \citep{walterscheid_reexamination_2003}.

\subsection{The instability growth rate}
\begin{figure} 
	\centerline{\includegraphics[width=0.65\textwidth]{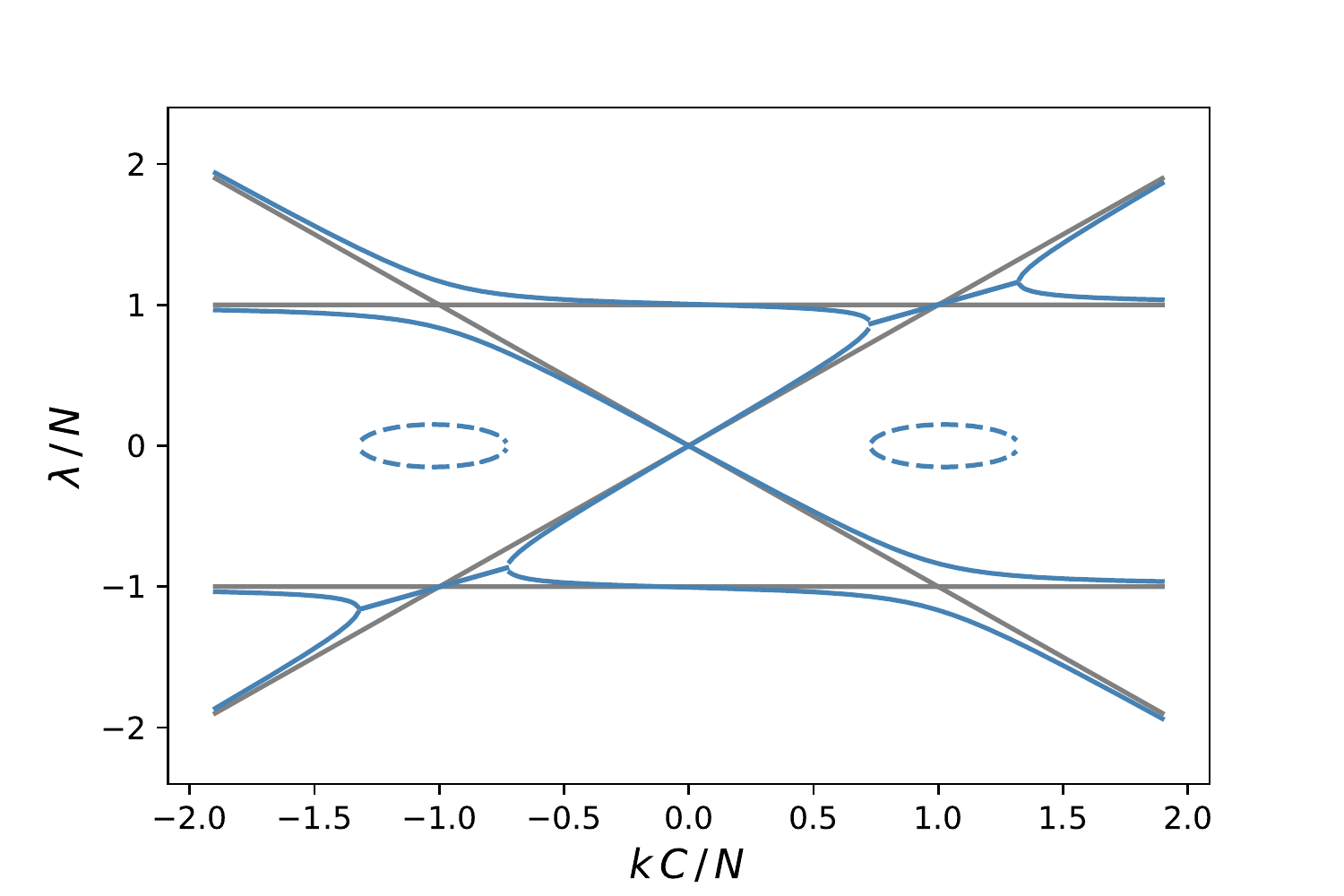}}
	\caption{Leading-order roots of the characteristic polynomial $\mathcal{P}_{\mathbf{S}(K,G;\,0)}\left(i\Lambda^{(0)}\right)$ (gray lines, see \eqref{eq:charpol}). Real part of the numerically computed roots of $\mathcal{P}_{\mathbf{S}(K,G;\,0.1)}(i\Lambda)$ (solid blue line) and its imaginary part (dashed blue line) for $\varepsilon=0.1$. We chose an unrealistically large $\varepsilon$ in order to exemplify its effect.}
	\label{fig:lead_dispersion}
\end{figure}

Figure \ref{fig:lead_dispersion} depicts all four leading-order roots (gray lines) for varying horizontal wavenumber $K$.
Notice that, to leading order, we obtain only real-valued roots, and these translate to purely imaginary and hence stable eigenvalues (spectrum). However, the numerical solution for $\varepsilon\neq 0$ reveals imaginary parts for $\Lambda$ (the dashed blue lines) and, consequentially, the perturbations might be unstable in these regions. Referring to Figure~\ref{fig:lead_dispersion}, we furthermore observe that the potentially unstable eigenvalues appear around $K=kC/N=\pm 1$ where the dispersion curves of Lamb and Brunt waves collide. This point is distinct as the leading-order operator degenerates: its eigenvalues have algebraic multiplicity of two instead of one at this point. \citet{perez_unidirectional_2022} identified and investigated almost the same points of degeneracy. 
Remarkably, they found a novel unidirectional mode with this approach.
The main difference to this study is induced by the dissimilar boundary conditions investigated here.

We now determine the maximum imaginary part of $\Lambda$, and this would give us the instability growth rate of the perturbation. To do so entails solving for an asymptotic solution at the point of degeneracy, $K=\pm 1$. However, perturbation theory for non-Hermitian operators that have leading-order eigenvalues with algebraic multiplicity greater than one is in fact cumbersome \citep{kato_short_1982}. Its rigorous mathematical treatment is beyond the scope of this paper as standard techniques from, say, quantum mechanics are inapplicable. That being the case, we will nevertheless provide an asymptotic analysis of the next-order correction to the eigenvalue, while refraining from the more involved computations for the eigenvectors. For this purpose, we expand $\Lambda$ in terms of powers of $\varepsilon$, i.e.,
\begin{align}
    \Lambda=\Lambda^{(0)}+\varepsilon^{1/2}\Lambda^{(1/2)}+\mathit{o}(\varepsilon^{1/2}).
\end{align}
The choice for the exponent of $1/2$ for the next-to-leading-order term is due to the multiplicity of the eigenvalue. We refer the interested reader to a complete reasoning on this matter in \citet{lidskii_perturbation_1966}. Also, \cite{kato_short_1982} contains a proof on the existence and form of the expansion. Substituting the ansatz together with the leading-order result in the characteristic polynomial \eqref{eq:charpol} and collecting terms of the same power in $\varepsilon$, we find that the $\mathit{O}(\varepsilon^{1/2})$ terms cancel each other. At $\mathit{O}(\varepsilon)$, we obtain%
\begin{align}
    \Lambda^{(1/2)}=\pm i\sqrt{\frac{G}{2}}.
    \label{eqn:lambda_half}
\end{align}
Rewriting \eqref{eqn:lambda_half} together with the leading-order result in a dimensional form yields
\begin{align}
    \lambda\approx\pm N\pm i\sqrt{\frac{F}{2}C\Gamma}
    \label{eqn:instability_growth}
\end{align}
and hence the instability growth rate close to the equator is given as $\sqrt{\Omega C\Gamma}$.

\subsection{Influence of the \texorpdfstring{$\beta$}{β}-plane effect}
\label{subsec:beta_influence}
We have assumed the $f$-$F$-planes up to this point of our theoretical investigations.
To conclude our theoretical investigations, we now discuss a possible influence by the $\beta$-plane effect
which may be estimated by means of scale analysis.
%We may use scale analysis for an estimate of the influence of the $\beta$-plane effect. 

Given an isothermal atmosphere such that $\bar{T}(z) \equiv T_0 = 300$\,K, $\gamma = 1.4$, $g \approx 9.81$\,m\,s$^{-1}$, 
$\Omega \equiv \Omega_y = 7.292 \times 10^{-5}$\,s$^{-1}$
and $R=287.4$\,J\,kg$^{-1}$\,K$^{-1}$, we obtain $N\approx 0.02$\,s$^{-1}$ and $C\approx 347$\,m\,s$^{-1}$.
With $k=N/C$, this leads to a wavelength of the unstable mode of approximately 120\,km, and a theoretical instability growth rate of
\begin{equation}
    \sqrt{\Omega C \Gamma} \approx 7.9 \times 10^{-4}\,\text{s}^{-1}.
    \label{eqn:theo_instability_growth}
\end{equation}
This instability growth rate corresponds to a doubling period of the perturbation's amplitude approximately every 15\,min.

By some Wentzel-Kramers-Brillouin-type argument, we may assume that the perturbation is meridionally homogeneous if it remains relatively constant within a latitude range of 10~wavelengths. Let us now compute the rate of the $\beta$-plane effect due to the term $f^\prime=\beta y$ at the northern edge of a meridional domain of size $10\times$120\,km~$=$~1200\,km. Taking $\beta=2\Omega/a$ at the equator where $a$ represents the Earth's radius, this amounts to $f'\approx 1\times 10^{-5}$\,s$^{-1}$.
As the $\beta$-plane effect is linear in the governing equations, we may compare it with the theoretical growth rate obtained from linear stability analysis and from \eqref{eqn:theo_instability_growth}, we observe that the theoretical growth rate is approximately 80 times greater than $f^\prime$. Hence, while damping of the perturbation might be possible due to the $\beta$-plane effect, the damping rate would be almost two orders of magnitude smaller than the instability growth rate. We therefore anticipate that the inclusion of the $\beta$-plane approximation will have only a {small} effect on the theory and results presented in this paper.

Summarising this section, we found a linearly unstable meridionally homogeneous mode of the hydrostatic atmosphere at rest close to the equator by means of asymptotic perturbation theory. The mode is to leading order identical to Lamb waves which are a known horizontally propagating external mode of the atmosphere. Its phase velocity is the speed of sound. 
To leading order, a free slip boundary condition was assumed which prohibits any vertical energy flux. We derived a higher-order correction to the eigenvalues of the linearised system and provided justifications that an analytical determination of the corresponding correction to the eigenvector is beyond the scope of this paper. Yet, in the next section we do obtain numerical approximations to the eigenmode structure and we use them to initiate simulations based on a full nonlinear compressible solver (see the discussion of \eqref{eqn:isotherm_eigenvalue_prob} and \eqref{eqn:initial_perturbation} below).

The validity of our asymptotic theory is therefore tested against the numerical solution of the nonlinear governing equations in the next section.

\section{Numerical Experiment and Results}
\label{sec:num_experiments}

\subsection{Numerical Model}
\label{sec:num_model}

Numerical corroboration of the Lamb-wave-like unstable mode that we derived in the previous section necessitates a model that is capable of solving the compressible non-hydrostatic Euler equations. The second-order accurate and semi-implicit numerical model by \cite{bk2019} is well-tested and is especially suitable to study flow instabilities numerically due to its conservation characteristics and numerical stability properties. A brief summary of the model is given in Appendix~\ref{apx:num_model}.
Details on the extension of the numerical model to support the non-traditional setting are documented in Appendix~\ref{apx:extension}.

\subsection{The initial condition}
\label{subsec:ics}
The initial condition for the numerical experiments is obtained by solving the eigenvalue problem \eqref{eqn:eigenvalue_problem} with the isothermal background assumption,
\begin{equation}
    (\mathbf{\tilde{T}} - i {\lambda})\tilde{\pmb{\psi}} = 0.
    \label{eqn:isotherm_eigenvalue_prob}
\end{equation}
This is done numerically, utilising the \textit{LAPACK} library's eigenvalue problem solver. The transformed perturbation variables $\pmb{\chi}$ are then obtained from the solution $\tilde{\pmb{\psi}}=(\tilde{\psi}_u,\,\tilde{\psi}_w,\,\tilde{\psi}_\theta,\,\tilde{\psi}_\pi)^\mathrm{T}$ by applying \eqref{eqn:soln_ansatz} and \eqref{eq:pert_ansatz}. Specifically, the initial perturbation fields are given as follows,
\begin{subequations}
\begin{align}
    u^\prime(x,z,0) &= A \tilde{\psi}_u \sqrt{\frac{\rho_0}{\bar{\rho}(z)}} \exp({-\Gamma z}) \, \cos\left( \frac{N}{C} x \right), \\
    w^\prime(x,z,0) &= A \tilde{\psi}_w \sqrt{\frac{\rho_0}{\bar{\rho}(z)}} \exp({-\Gamma z}) \, \cos\left( \frac{N}{C} x \right), \\
    \theta^\prime(x,z,0) &= A \tilde{\psi}_\theta \frac{N}{g} \sqrt{\frac{\rho_0}{\bar{\rho}(z)}} \bar{\theta} \exp({-\Gamma z}) \, \cos\left( \frac{N}{C} x \right), \\
    \pi^\prime(x,z,0) &= A \tilde{\psi}_\pi \frac{C}{c_p} \sqrt{\frac{\rho_0}{\bar{\rho}(z)}} \bar{\theta}^{-1}(z)\exp({-\Gamma z}) \, \cos\left(\frac{N}{C} x \right),
\end{align}%
\label{eqn:initial_perturbation}%
\end{subequations}%
where $A = 10^{-1}$\,m\,s$^{-1}$ is a prescribed small amplitude. Furthermore, the background density is given as
\begin{equation}
    \bar{\rho}(z) = \rho_0 \exp(-z / H_{\rho}),
\end{equation}
with $H_{\rho} = RT_0 / g$ representing the density scale height. The expression for $\Gamma$ in \eqref{eqn:Gamma} simplifies to
\begin{equation}
    \Gamma = \frac{1}{H_{\rho}} \left( \frac{1}{\gamma} - \frac{1}{2} \right)
\end{equation}
under the assumption of an isothermal atmosphere, and the background potential temperature is
\begin{equation}
    \bar{\theta}(z) = \theta_0 \exp(z/H_{\theta}),
\end{equation}
where $H_{\theta} = H_{\rho} / \kappa$ denotes the potential temperature scale height. The quantities $\rho_0$, $T_0$, and $\theta_0$ are reference density, temperature, and potential temperature at $z=z_0$ respectively. The initial density field is obtained via the equation of state \eqref{eq:ideal}.

The horizontal extent of the domain is chosen such that it is equivalent to four wavelengths, i.e., ${x_\mathrm{MAX}} = 2\pi j C / N$ with $j=4$, and this corresponds to $x \in [-244.5\,\mathrm{km},\,244.5\,\mathrm{km}]$. The full vertical extent of the domain corresponds to $z\in[0.0,\,80.0\,\mathrm{km}]$. 

\subsection{The total energy flux}
\begin{figure}[h]
	\centerline{\includegraphics[width=0.55\textwidth]{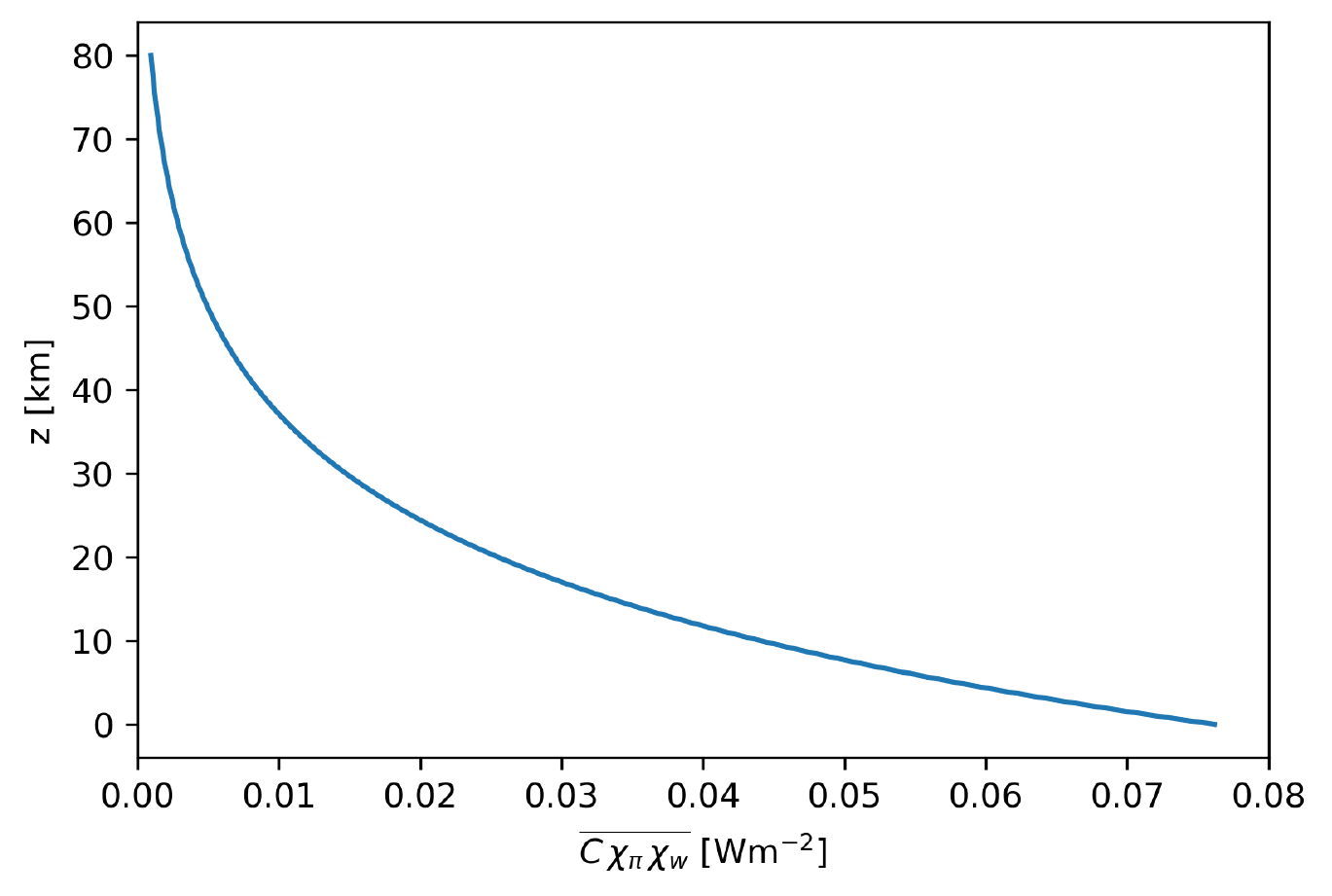}}
	\caption{Initial horizontally-averaged vertical flux of the total energy.}
	\label{fig:vertical_energy_flux}
\end{figure}

From the initial condition, we additionally computed the vertical flux of total energy as given in \eqref{eq:roc_glob_energy}. 
Figure~\ref{fig:vertical_energy_flux} shows the profile of the horizontal mean of the vertical energy flux.
The positive vertical energy flux indicates an upward flow of energy and decays quickly for increasing $z$.
In line with physical expectations and as discussed in Subsection~\ref{subsec:linear_analysis}, energy enters the domain through the bottom boundary but cannot escape into space, resulting in an accumulation of energy in the domain and generating a growing instability.
Notice that the initial flux at the bottom boundary is very small, as expected from the asymptotic theory. 
This initial flux is also small compared to typical sensible or latent heat fluxes in the atmosphere especially considering the initial amplitude of 0.1\,ms$^{-1}$, which presents a recognisable breeze.
Since our simulation domain is finite and the flux at the top is non-zero, an absorbing non-reflecting boundary condition at the TOA is necessary.

\subsection{Numerical representation of the boundary conditions}
To emulate a free surface at the top of the boundary, a wave-absorbing layer is applied to the top 20\,km to approximate the non-reflecting boundary condition \citep{durran2010numerical}. Specifically, the Rayleigh damping by \citet{durran1983compressible}, as reproduced below, has been used. The following terms are added to the right-hand side of the governing equations in \eqref{eq:govern},
\begin{subequations}
\begin{align}
    R_u &= \tau(z) u^\prime,\\
    R_w &= \tau(z) w^\prime,\\
    R_\theta &= \tau(z) \theta^\prime,\\
    R_\pi &= \tau(z) \pi^{\prime},
\end{align}
\end{subequations}
with
\begin{equation}
    \tau(z) = 
    \begin{cases}
    0 & \text{for}~z\leq z_D, \\
    -\frac{\alpha}{2} \left[ 1 - \cos \left(\frac{z-z_D}{z_{\rm TOA} - z_D} \pi \right)  \right] & \text{for}~0 \leq \frac{z-z_D}{z_{\rm TOA} - z_D} \leq \frac{1}{2}, \\ 
    -\frac{\alpha}{2} \left[ 1 + \left( \frac{z-z_D}{z_{\rm TOA}-z_D} - \frac{1}{2} \right) \pi \right] & \text{for}~\frac{1}{2} \leq \frac{z-z_D}{z_{\rm TOA} - z_D} \leq 1,
    \end{cases}%
\end{equation}%
where $z_{\rm TOA}=80.0$\,km, $z_D=60.0$\,km, and $\alpha=0.5$. An inverse of the Rayleigh damping is applied to the bottom 3\,km of the domain. Instead of relaxing the flow to the ambient state, this bottom \textit{forcing} relaxes fully to the perturbation variables obtained from the numerical solution of the eigenvalue problem in \eqref{eqn:initial_perturbation}, which will be referred to as the ``semi-analytical eigenmode forcing'' (SA) below. Such a bottom forcing mimics an atmospheric boundary layer that allows a vertical energy transport into the free atmosphere.

Both the top and bottom boundary conditions are approximations of the physical boundaries of the atmosphere, and these may be sources of error. See Subsection~\ref{subsec:linear_analysis} for more details on these choices of the top and bottom boundaries. 

Furthermore, the domain is periodic in the horizontal direction. The simulations in this section are conducted on a grid with $(N_x \times N_z)= (301 \times 120)$ grid points.

\begin{figure}[!h]
    \centerline{\includegraphics[width=1.0\textwidth]{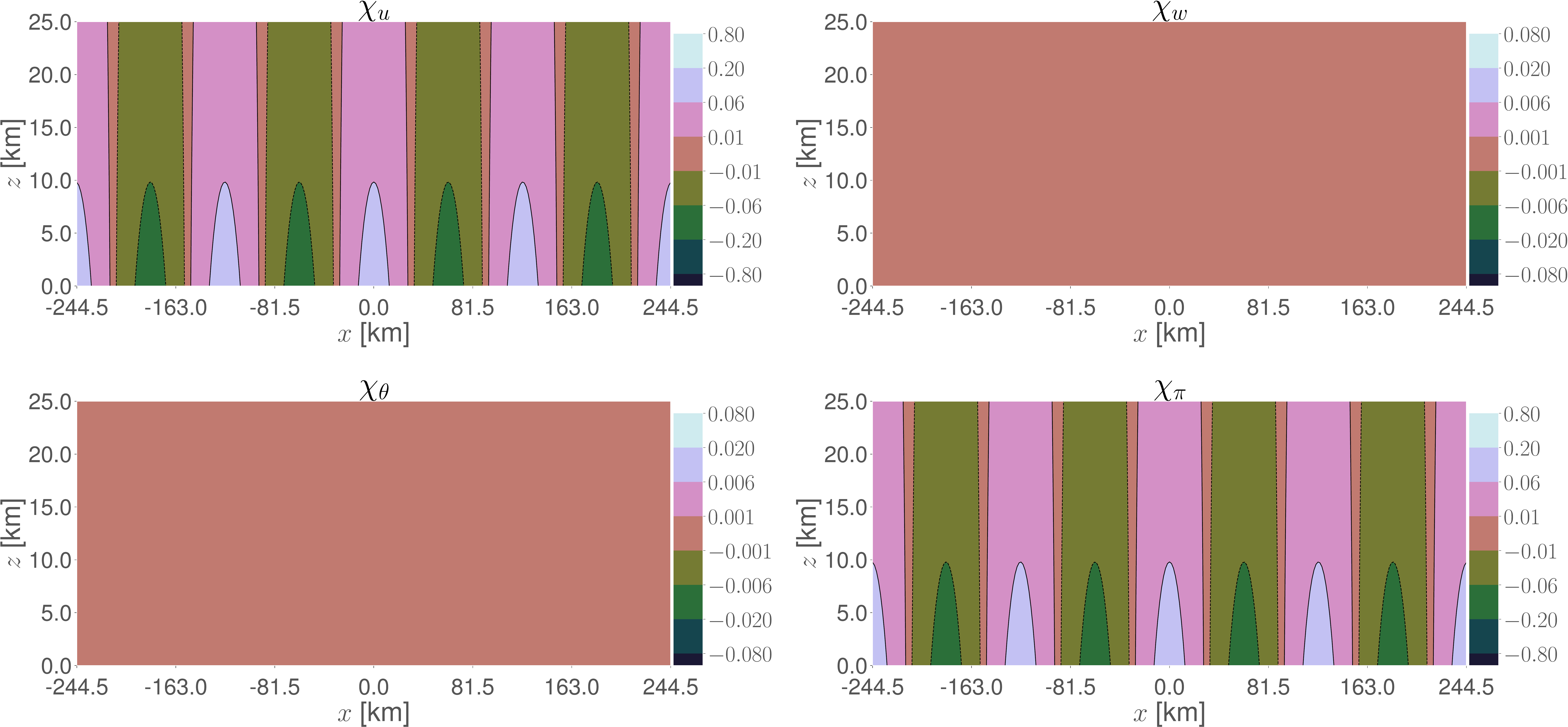}}
    \caption{Initial perturbation: distribution of the prognostic variables with the transformation given in \eqref{eqn:hat_eqns} for the stable Lamb wave (LW) case. (top left) Transformed horizontal velocity perturbation $\chi_u$, (top right) transformed vertical velocity perturbation $\chi_w$, (bottom left) transformed potential temperature perturbation $\chi_\theta$, and (bottom right) transformed Exner pressure perturbation $\chi_\pi$. The contours have the unit of kg$^{1/2}$\,m$^{-1/2}$\,s$^{-1}$, i.e., the units of the square root of energy density.}
    \label{fig:trad_ICs}
\end{figure}

\begin{figure}[!h]
    \centerline{\includegraphics[width=1.0\textwidth]{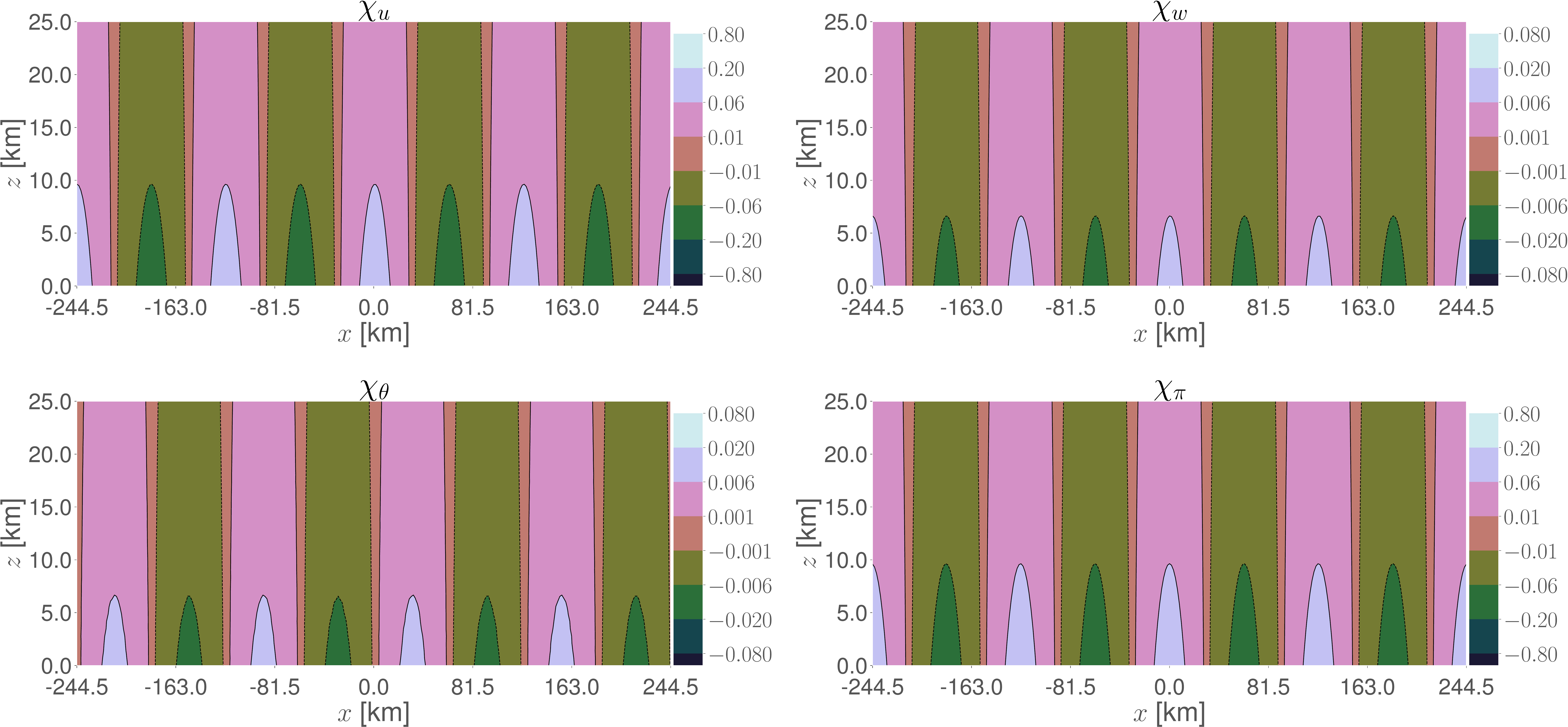}}
    \caption{Output of the initial transformed prognostic variable fields for the Lamb-wave-like instability depicting the initial non-traditional Coriolis effect, i.e., $\Omega_y = 7.292 \times 10^{-5}$\,s$^{-1}$ at $t=0$\,s, corresponding to the initial conditions of the LWLI-SA and LWLI-SO cases. Contour units are as in Figure~\ref{fig:trad_ICs}.}
    \label{fig:unstable_ICs}
\end{figure}

For the initial perturbation in \eqref{eqn:initial_perturbation}, the prognostic variables in the form of \eqref{eqn:hat_eqns} are depicted in Figure~\ref{fig:trad_ICs} for the Lamb wave and Figure~\ref{fig:unstable_ICs} for the unstable mode. Note that the colour scale employed in these and the subsequent contour plot Figures are nonlinear.

Two further points must be addressed before we present our results in the subsequent Subsection. In Appendix~\ref{apx:lamb_bal}, we present a careful numerical treatment of the thermodynamical quantities that ensures a well-balanced Lamb wave solution in a stably stratified atmosphere. With this development, the linear solution of the Lamb wave in the traditional setting maintains a vertical velocity field up to machine accuracy for 10\,000 time steps. In Appendix~\ref{apx:energy}, we investigate the energy conservation of the stable Lamb wave solutions with the numerical scheme. In particular, we demonstrate that energy is conserved up to $\sim0.4\%$ over 36\,000\,s for a run with Lamb wave initial condition in the traditional setting, i.e., with $F=0$\,s$^{-1}$ in \eqref{eqn:isotherm_eigenvalue_prob}, and top and bottom rigid wall boundaries. This is in line with the theory developed in Section~\ref{sec:theory}. Furthermore, energy is similarly well -- though not exactly -- conserved for a run that is initialised with a stable Lamb wave but features non-zero Coriolis strength.

\subsection{Instability growth rate under the non-traditional setting}
\label{subsec:instab_growth}
Three experimental configurations are investigated below: a run with the Lamb-wave-like unstable eigenmode as the initial condition, with Coriolis acceleration in the non-traditional setting, and the semi-analytical eigenmode bottom forcing (LWLI-SA); a run with the Lamb-wave-like unstable eigenmode as the initial condition, with Coriolis acceleration in the non-traditional setting, and sub-optimal bottom forcing (LWLI-SO--an explanation on the sub-optimal forcing is given below); and a stable Lamb wave as the initial condition without Coriolis acceleration and bottom forcing (LW).
The time-step size $\Delta t=10$\,s is restricted by an advective Courant-Friedrichs-Lewy number of $\rm{CFL}_{\rm adv}=0.9$.
Having established the stability of the Lamb wave with the traditional approximation in Appendix~\ref{apx:energy} and in particular Figure~\ref{fig:stabGrowth}, the focus of our study below will be on the Lamb-wave-like instability in the non-traditional setting with a Rayleigh damping at the top and a bottom forcing corresponding to the semi-analytical eigenmode that can be obtained by the numerical solution to the eigenvalue problem.

\begin{figure}[h]
    \centerline{\includegraphics[width=1.0\textwidth]{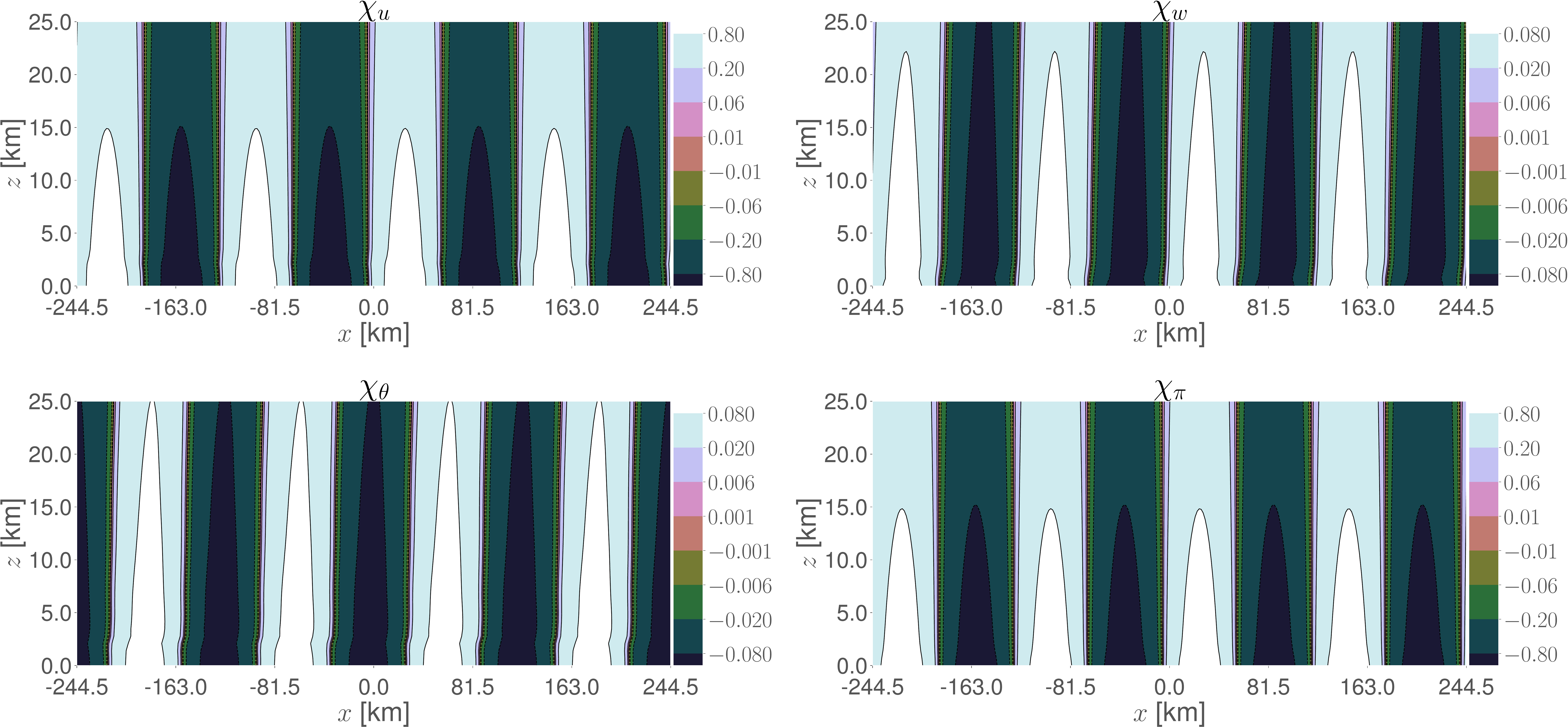}}
    \caption{Output of the fields for the transformed prognostic variables at $t=1$\,hrs for the LWLI-SA case, i.e., a Lamb-wave-like instability in the non-traditional setting and the semi-analytical bottom forcing. Contour units are as in Figure~\ref{fig:trad_ICs}.}
    \label{fig:instabFinal}
\end{figure}

Figure~\ref{fig:instabFinal} depicts the fields at $3\,600$\,s$~=1$\,hr for the LWLI-SA case with the non-traditional setting.
Here we observe a significant increase of the amplitude of the perturbation due to the instability. The structure and amplitude of these fields are similar to the corresponding solution of the eigenvalue problem in \eqref{eqn:isotherm_eigenvalue_prob}, i.e., at $t=1$\,hr and $F=2\Omega_y=1.458\times10^{-4}$\,s$^{-1}$. These results therefore closely match the structure of the theoretically predicted instability from Section~\ref{sec:theory}.

\begin{figure}[!h]
    \centerline{\includegraphics[width=1.0\textwidth]{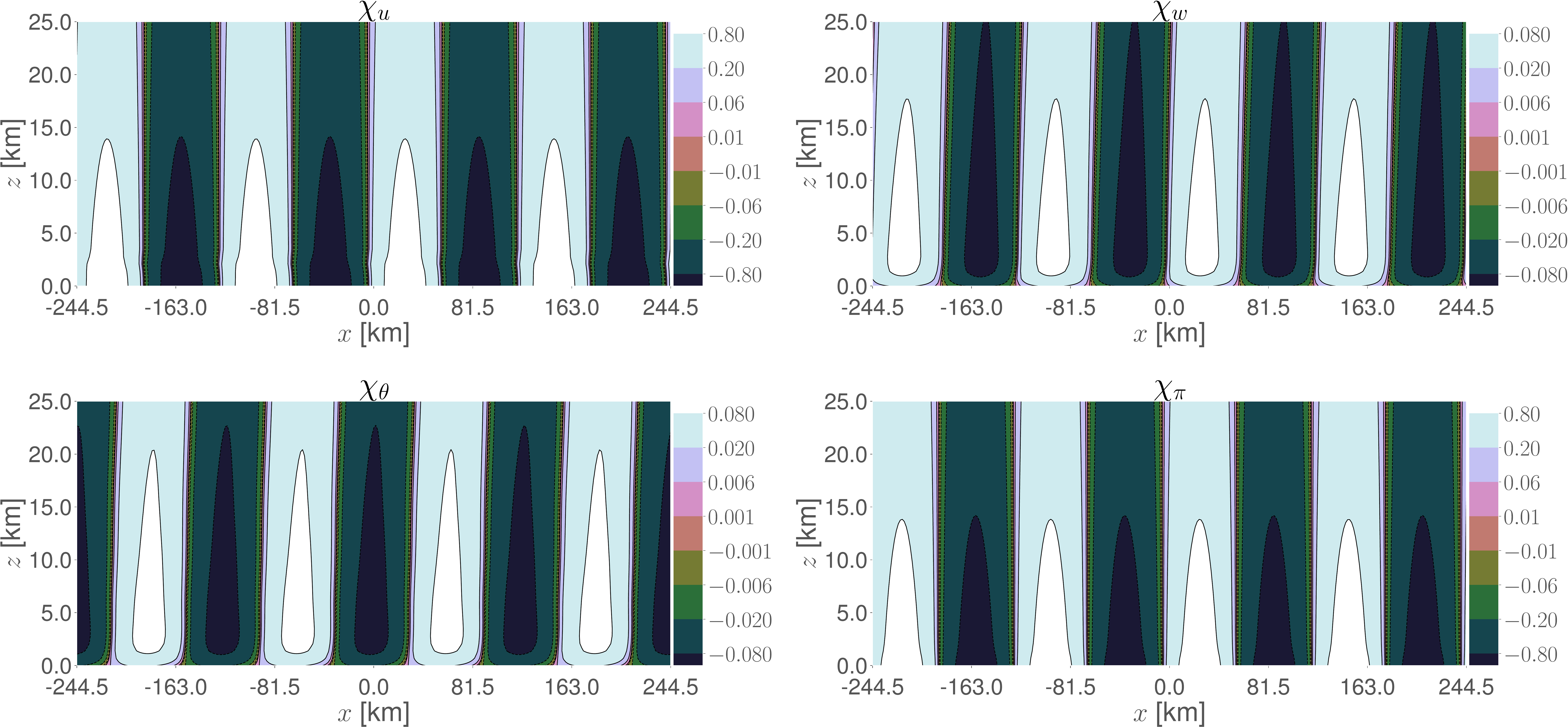}}
    \caption{Output of the fields for the transformed prognostic variables at $t=1$\,hrs for the LWLI-SO case, i.e., a Lamb-wave-like instability in the non-traditional setting and sub-optimal bottom forcing. Contour units are as in Figure~\ref{fig:trad_ICs}.}
    \label{fig:subopFinal}
\end{figure}

For completeness, we include the LWLI-SO result with a sub-optimal bottom forcing in Figure~\ref{fig:subopFinal}. This is done by only applying a forcing to the quantities $u^\prime$ and $\pi^\prime$. This test aims to mimic more physical, real-world scenarios where there might be an injection of energy from the bottom atmospheric boundary layer, but that may not necessarily correspond to the non-traditional semi-analytical eigenmode structure. Outside of the bottom 3\,km wherein the forcing is applied, the solution structure after 1\,hr looks almost identical to the structure of the instability in Figure~\ref{fig:instabFinal}. The results indicate that a similar instability develops even if the bottom forcing deviates from the optimal structure.

In order to quantify our observations of the simulation results, 
we compute the relative norm as a measure of the energy of the perturbation. The relative norm is given by
\begin{equation}
    \text{rel. norm} = \frac{\Vert {\pmb{\chi}} \Vert}{\Vert{\pmb{\chi}}_0 \Vert},
\end{equation}
where $\Vert \, \cdot \, \Vert$ represents the $L^2$-norm in \eqref{eqn:2-norm}, and $\pmb{{\chi}}_0$ denotes the initial $\pmb{{\chi}}$.

\begin{figure}[!h]
    \centerline{\includegraphics[width=0.55\textwidth]{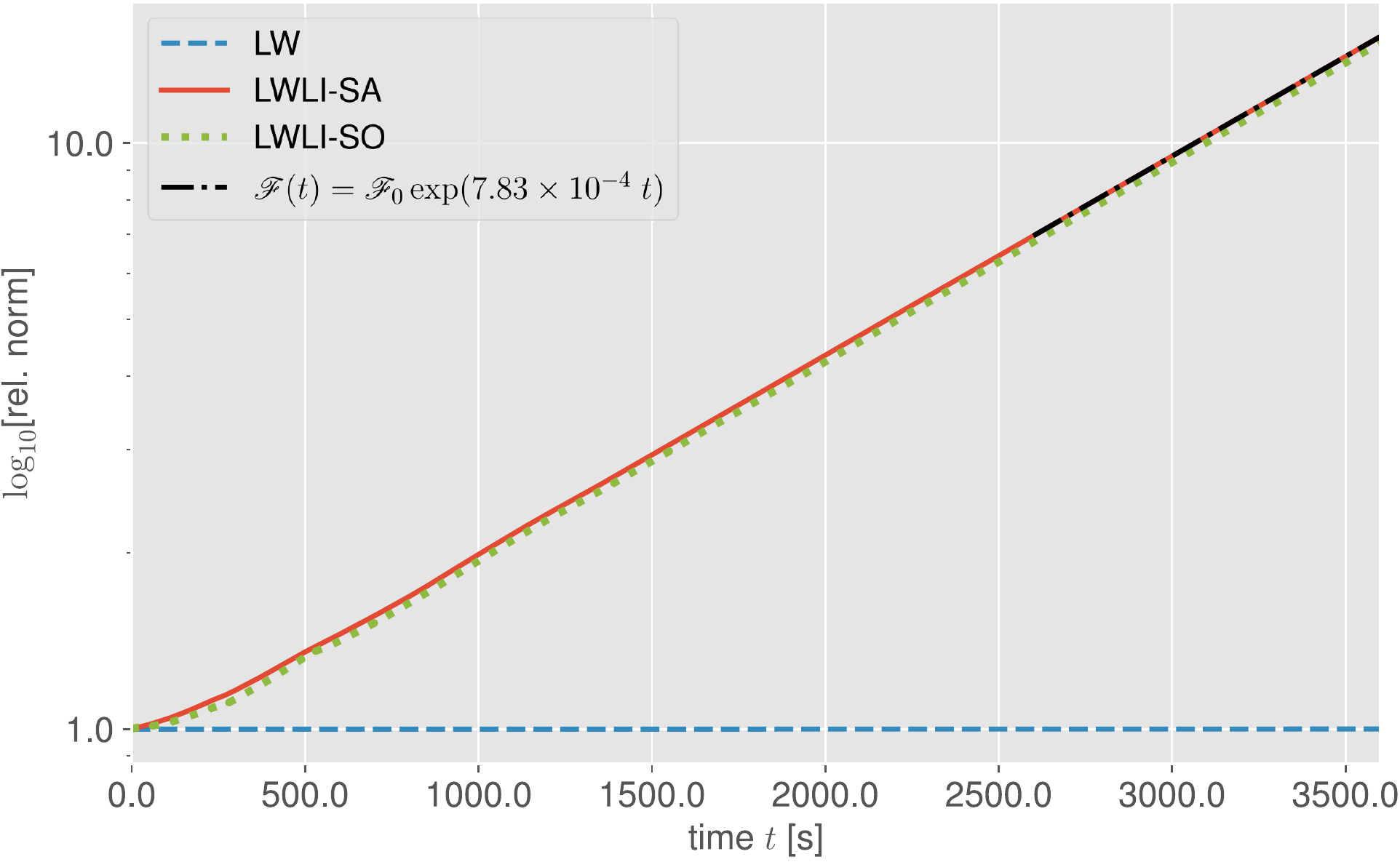}}
    \caption{Semi-logarithmic plot of the relative norm against time: a run with stable Lamb wave initial conditions in the traditional approximation (LW; blue dashed curve); a run with unstable Lamb-wave-like initial conditions in the non-traditional setting and semi-analytical eigenmode as the bottom forcing (LWLI-SA; red solid curve); a run with the unstable Lamb-wave-like initial conditions in the non-traditional setting and sub-optimal bottom forcing (LWLI-SO; green dotted curve); and a best-fit curve (dash-dotted in black) of the increment in the LWLI-SA run. The relative norm for all runs is taken in the limited $z\in[3.0,\,25.0]$\,km domain.}
    \label{fig:instabGrowth}
\end{figure}

Figure~\ref{fig:instabGrowth} depicts the logarithm of the relative norm over time for the three configurations. The relative norm is computed over the vertical domain of $z \in [3.0, 25.0]$\,km for all three runs, i.e., we effectively exclude the lowest 3km from our analysis. This is to exclude energy contributions from the bottom forcing.
As demonstrated in Appendix~\ref{apx:energy}, the relative norm of the traditional approximation LW run (blue dashed curve) remains close to the initial value, indicating that the energy of the system is almost exactly conserved over the time scale considered here. For the non-traditional setting LWLI-SA run with the semi-analytical eigenmode structure applied as the bottom forcing (red curve), a best-fit curve is plotted to quantify the instability growth rate (dash-dotted curve in black). Specifically, the best-fit curve is given by
\begin{equation}
    \mathscr{F}(t) = \mathscr{F}_0 \exp({\zeta}\,t),
    \label{eqn:best_fit}
\end{equation}
where $\mathscr{F}_0$ and $\zeta$ are obtained from the curve-fitting method. From this best-fit curve, we find an instability growth rate of about $7.83 \times 10^{-4}$\,s$^{-1}$, which deviates from the theoretically predicted value in \eqref{eqn:theo_instability_growth} by merely $0.8\%$. Finally, a similar relative norm curve for the non-traditional setting LWLI-SO run with the sub-optimal bottom forcing is depicted (dotted green curve), and the instability growth rate is equally close to the theoretical value as it yields the same growth rate of $7.83 \times 10^{-4}$\,s$^{-1}$ up to the third digit (best-fitted curve is not shown here).

All experiments presented in this section were performed with the same spatial and temporal resolutions. However, additional runs were conducted with spatial resolutions of $(151 \times 60)$ and $(301 \times 120)$. Furthermore, for each spatial grid resolution, runs with the following time-step sizes $\Delta t$ were carried out (results apart from the configuration studied above are not shown):
\begin{equation}
    \Delta t = (1, 2, 4, 8, 10, 12, 14, 16)\,\text{s}.
    \label{eqn:varying_time_res}
\end{equation}
With values in the range of $(7.83\pm0.01) \times 10^{-4}$\,s$^{-1}$, the growth rates obtained from all of these runs do not differ significantly from those presented above. The task of estimating the instability growth rate is difficult and may be sensitive to the experimental parameters \citep{mcnally2012well,zhelyazkov2015modeling}. Therefore, the consistent experimental growth rate obtained here across varying temporal and spatial resolutions strongly implies a dynamical nature of the unstable mode.

The numerical results presented in this Section resolutely suggest that the unstable mode described in Section~\ref{sec:theory} may be expected to arise in atmospheric flow simulations that implement the non-traditional Coriolis acceleration. In particular, any problem formulation with a boundary layer model that does not per design suppress an energy influx into the bulk of the troposphere from below will permit these growing modes.

\section{Conclusions}
\label{sec:conclusions}
% short statement on the impact of the result of this paper
This paper presents a novel unstable mode that arises from considering the non-traditional terms of the Coriolis force. This unstable mode lends itself to physical interpretations and may have substantial implications for numerical schemes that employ the non-traditional setting.

% Summary of the paper: theory
Considering the linearised compressible Euler equations in the non-traditional setting, an inner product of the state vector of the prognostic variables for perturbations is derived inducing a norm. We showed that this norm is proportional to the perturbation's total energy. We also found that the differential operator emerging from the linear system, after the usual separation of variables, is not skew-Hermitian with respect to the inner product when the boundaries are non-trivial. By some version of the spectral theorem, this observation gives rise to a potentially unstable spectrum. 
And indeed, by means of asymptotic theory, we showed that the isothermal, hydrostatic atmosphere at rest subject to the non-traditional Coriolis acceleration becomes prone to an unstable Lamb-wave-like mode 
under infinitesimal initial perturbations.
The theoretical growth rate of the unstable mode is quantified.

% Summary of the paper: numerics
With a small perturbation of the isothermal hydrostatic atmosphere at rest, numerical experiments demonstrate that this initial configuration supports a linearly unstable mode when the full Coriolis acceleration is considered. The experimental instability growth rate is close to the theoretical value. In contrast, the simulation remains stable if the traditional approximation is applied. 
Even if the injection of energy at bottom boundary is sub-optimal, i.e., if it deviates from the theoretical structure, a similar unstable mode develops nevertheless. Therefore, this unstable mode may present itself under conditions that are less stringent than the theory implies.

% physical implications
Physical evidence from observations for this novel type of unstable mode can be found in \cite{nishida_background_2014}. They observed Lamb waves at around $30^\circ$~North with a vast array of high-resolution microbarometers. In contrast to intermittent waves due to large-scale events such as volcanic eruptions, they found omnipresent background waves. The authors discussed several excitation mechanisms as a cause for the phenomenon observed. The results presented in this paper imply that these Lamb waves might be generated by the instability due to the Coriolis force, as the frequency spectrum of the observations fits well with our theoretical predictions. Further corroboration is that most waves the authors observed travel along the East-West axis as predicted by our theory.

% numerics implications
Furthermore, our investigation presents relevant implications for numerical models that use the non-traditional setting. One would encounter this unstable mode in numerical simulations if there is energy flux from the atmospheric boundary layer. This effect would be especially significant for the numerical simulations of the deep atmosphere, e.g. the papers by \citet{borchert_upper-atmosphere_2019,smolarkiewicz_finite-volume_2016} as mentioned in the introduction, or in tropical dynamics with the non-traditional setting. Therefore, the characterisation of the unstable mode presented here is a worthwhile pursuit that may broaden our understanding of the simulation results from these numerical models.

% limitations and outlook
Our derivation of the unstable mode and the numerical experiments have two limitations that may be investigated in a future study. 1) The effect of the Earth's curvature has not been considered, and 2) meridional homogeneity was assumed. On point 1), it is possible to account for the Earth's curvature by incorporating its effect in the $\beta$-plane term following \citep{maas2007equatorial}. However, in contrast to these authors, we have considered the compressible rather than the incompressible Euler equations in this study, adapting their approach to the present application will require substantial work, which we leave to a potential future study. On point 2), we plan to carry out 3D numerical experiments that will explore the influence of the $\beta$-plane effect on the eigenmodes found in the present study. An exhaustive examination of the physical interpretations of the unstable mode could also be carried out. 

Finally, it may also be fruitful to investigate the unstable mode in the more operationally-ready numerical models at the German Weather Service (DWD) and the European Centre for Medium-Range Weather Forecasts (ECMWF). Two questions are particularly worth answering. Can the unstable mode described in this paper be reproduced by the numerical models? If the answer is yes, what influences does the unstable mode have on atmospheric dynamics and circulation? 

\newpage

\section*{Acknowledgements}
{
The authors express their gratitude to the three anonymous reviewers for taking the time and effort to review this paper. The authors also thank Ulrich Achatz for the discussions that helped improve this manuscript.
}

\section*{Funding}
{
% Please provide details of the sources of financial support for all authors, including grant numbers. Where no specific funding has been provided for research, please provide the following statement: "This research received no specific grant from any funding agency, commercial or not-for-profit sectors." 
The authors thank the Deutsche Forschungsgemeinschaft for the funding through the Collaborative Research Center (CRC) 1114 ``Scaling cascades in complex systems'', Project Number 235221301, Project A02: ``Multiscale data and asymptotic model assimilation for atmospheric flows'' and through Grant KL 611/25-2 of the Research Unit FOR 1898 ``Multi-Scale Dynamics of Gravity Waves''. R.C. is furthermore grateful for the generosity of Eric and Wendy Schmidt through the Schmidt Futures VESRI ``DataWave'' project.
}

%\section*{Declaration of interests}
%{
%The authors report no conflict of interest.
%}

\section*{Data availability statement}
{Supplementary results similar to Figures \ref{fig:trad_ICs}--\ref{fig:instabGrowth} for the varying spatial and temporal resolutions mentioned in Section~\ref{sec:num_experiments} [cf. \eqref{eqn:varying_time_res}] are available at \url{https://doi.org/10.5281/zenodo.8010527}. The datasets used to generate the numerical results in this manuscript and the supplementary results are also available here.}

%\section*{Author ORCIDs}
%{
%% Authors may include the ORCID identifers as follows.  F. Smith, https://orcid.org/0000-0001-2345-6789; B. Jones, https://orcid.org/0000-0009-8765-4321
%R. Chew, \url{https://orcid.org/0000-0001-6454-8401}; M. Schlutow, \url{https://orcid.org/0000-0002-3640-634X}; R. Klein, \url{https://orcid.org/0000-0001-8032-3851}
%}

\section*{Author contributions}
{
M.S. developed the theoretical stability analysis that forms the basis of this work. R.C. and R.K. extended the numerical model to support the non-traditional setting and the well-balanced Lamb wave solution. R.C. also ran the numerical experiments and took the lead in writing the manuscript.
}

\newpage

\appendix
\section{The Coriolis force}
\label{apx:coriolis}
\begin{figure}[!h]
    \centering
    \begin{tikzpicture}
    \tikzstyle{every node}=[font=\small];
    
    \def\r{4cm};
    \def\ahs{2mm};

    \shade[ball color=gray!10, opacity = 0.25, domain=0:90] (0,0) -- plot ({\r*cos(\x)}, {\r*sin(\x)});
    
    \draw[domain=0:90] plot ({\r*cos(\x)}, {\r*sin(\x)});

    \draw[-{Latex[length=\ahs]}] (0,0) -- (0, \r+\r/5) node[pos=0.9] (axisrot) {\AxisRotator[x=0.1cm, y=0.4cm, rotate=90, line width=.08ex]} node[pos=1.0, right] {north pole};
    
    \draw[dashed] (0,0) -- (\r,0) node[pos=1.01, right] {equator};
    
    \node[right=-0.2cm of axisrot] {$\pmb{\Omega}$};
    
    \draw[-{Latex[length=\ahs]}] (0,0) -- ({\r*cos(35)}, {\r*sin(35)});
    
    \draw[-{Latex[length=\ahs]}, domain=0:35] plot ({(\r/4)*cos(\x)}, {(\r/4)*sin(\x)});
    
    \node at ({(\r/4)*cos(35/2)+\r/16}, {(\r/4)*sin(35/2)+\r/48}) {$\Phi$};
    
    \draw[-{Latex[length=\ahs]}, thick] ({\r*cos(35)}, {\r*sin(35)}) -- ({\r*cos(35)}, {\r*sin(35)+2cm}) node[pos=1.0, right]{$2\Omega$} ;
    
    \draw[-{Latex[length=\ahs]}, dashed, thick] ({\r*cos(35)}, {\r*sin(35)}) -- ({\r*cos(35)+(2cm*cos(35)*sin(35)}, {\r*sin(35)+(2cm*sin(35)*sin(35)}) node[pos=1.0, right]{$2\Omega \sin(\Phi)$};
    
    \draw[-{Latex[length=\ahs]}, dashed, thick] ({\r*cos(35)}, {\r*sin(35)}) -- ({\r*cos(35)-(2cm*cos(35)*sin(35)}, {\r*sin(35)+(2cm*cos(35)*cos(35)}) node[pos=1.2]{$2\Omega \cos(\Phi)$};
    
    \end{tikzpicture}
    \caption{An illustration of the effect of the full Coriolis acceleration acting on the Earth's surface at a latitude $\Phi$. An explanation is provided in the accompanying text.}
    \label{fig:coriolis}
\end{figure}
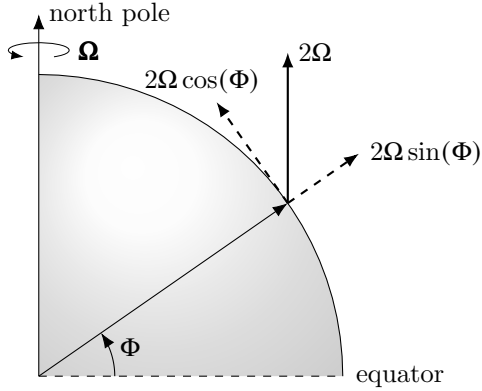

At a certain latitude represented by $\Phi$, the full Coriolis force on the Earth's surface is given by a titled vector with magnitude $2\Omega$. This may be written as a sine term perpendicular to the surface, and a cosine term that is parallel. The traditional approximation ignores contributions from the cosine terms.

\citet{thuburn_normal_2002} introduced the $f$-$F$-planes approximation that is analogous to the non-traditional approximation. Here, $f=2\Omega \sin(\Phi)$ and $F=2\Omega \cos(\Phi)$. The $\beta$-plane approximation extends the $f$-plane by including its latitudinal dependence, i.e.,
\begin{equation}
    f_\beta = f+ f^\prime = f + \beta y,
\end{equation}
with $\beta = (\mathrm{d}f / \mathrm{d}y)|_\Phi = 2\Omega \cos(\Phi) / a$, where $y$ is the meridional distance from the equator, and $a$ is the Earth's radius. In this paper, we consider the atmosphere close to the equator with the $f$-plane approximation only.

\section{The numerical model}
\label{apx:num_model}
In this Appendix, we describe the main structural features of the discretisation following \citet{bok2014,bk2019}. A summary of the numerical model is also provided in \citet{chew_one-step_2022}.

The governing adiabatic compressible equations with constant heat capacities under the influence of gravity and in a rotating coordinate system corresponding to a non-traditional tangent-plane approximation are given by
\begin{subequations}
    \begin{align}
        \frac{\partial \rho}{\partial t} + \nabla_\Vert \cdot (\rho \pmb{u}) + \frac{\partial (\rho w)}{\partial z} &= 0, 
        \label{eqn:num_gov_eqns_a} \\
        \frac{\partial (\rho \pmb{u})}{\partial t} + \nabla_\Vert \cdot (\rho \pmb{u} \circ \pmb{u}) + \frac{\partial (\rho w \pmb{u})}{\partial z} &= - \left[ c_p P \nabla_\Vert \pi + (2\,\pmb{\Omega} \times \rho \pmb{v})_{\Vert} \right], 
        \label{eqn:num_gov_eqns_b} \\
        \frac{\partial (\rho w)}{\partial t} + \nabla_\Vert \cdot (\rho \pmb{u}w) + \frac{\partial (\rho w^2)}{\partial z} &= - \left[ c_p P \, \frac{\partial \pi}{\partial z} + \rho g + (2\,\pmb{\Omega} \times \rho \pmb{v})_{\perp} \right] , 
        \label{eqn:num_gov_eqns_c} \\
        \frac{\partial P}{\partial t} + \nabla_\Vert \cdot (P \pmb{u}) + \frac{\partial (Pw)}{\partial z} &= 0.
        \label{eqn:num_gov_eqns_d}
    \end{align}
    \label{eqn:num_gov_eqns}
\end{subequations}
Here, $\pmb{u} = u \pmb{e}_x + v \pmb{e}_y$ is the velocity vector of the zonal and meridional wind, and
\begin{equation}
    P = \frac{p_0}{R} \left( \frac{p}{p_0} \right)^{\frac{c_v}{c_p}} \equiv \rho \theta
    \label{eqn:P_eos}
\end{equation}
is the mass-weighted potential temperature. The operator $\nabla_\Vert = \pmb{e}_x\,\partial/\partial x+\pmb{e}_y\, \partial /\partial y$ subsumes the horizontal partial derivatives, and $\pmb{q}_{\Vert},\,\pmb{q}_{\perp}$ are the horizontal and vertical components of the vector $\pmb{q}$.

Multiplying the governing equations \eqref{eq:govern} with $\rho$ leads to \eqref{eqn:num_gov_eqns}, wherein the evolution equation of the Exner pressure $\pi$ in \eqref{eq:govern_c} is now subsumed under the evolution equation for $P$, i.e., \eqref{eqn:num_gov_eqns_d}. This particular formulation of the governing equations in \eqref{eqn:num_gov_eqns} facilitates the subsequent discussion on the numerical model.

\subsection{Compact description of the time integration scheme}

\subsubsection{Reformulation of the governing equations}
The primary unknowns advanced in time are the same as in \eqref{eqn:num_gov_eqns}, i.e., $(\rho, \rho\pmb{u}, \rho w, P)$. Moreover, the inverse of the potential temperature is introduced as
\begin{equation}
    \chi = 1 / \theta,
\end{equation}
such that \eqref{eqn:num_gov_eqns_a} becomes a transport equation for $\chi$,
\begin{equation}
    \frac{\partial \rho}{\partial t} + \nabla_\Vert \cdot (\rho \pmb{u}) + \frac{\partial (\rho w)}{\partial z} = \frac{\partial (P\chi)}{\partial t} + \nabla_\Vert \cdot (P\chi \pmb{u}) + \frac{\partial (P\chi w)}{\partial z} = 0,
\end{equation}
with $(P\pmb{v})$ the advecting flux. The governing equations now read as
\begin{subequations}
    \begin{align}
        \frac{\partial \rho}{\partial t} + \nabla_\Vert \cdot (P \pmb{u} \chi) + \frac{\partial (P w \chi)}{\partial z} &= 0,
        \label{eqn:reform_num_gov_eqns_a} \\
        \frac{\partial (\rho \pmb{u})}{\partial t} + \nabla_\Vert \cdot (P \pmb{u} \circ \chi \pmb{u}) + \frac{\partial (P w \chi \pmb{u})}{\partial z} &= - \left[ c_p P \nabla_\Vert \pi + (2\,\pmb{\Omega} \times \rho \pmb{v})_{\Vert} \right], 
        \label{eqn:reform_num_gov_eqns_b} \\
        \frac{\partial (\rho w)}{\partial t} + \nabla_\Vert \cdot (P \pmb{u} \chi w) + \frac{\partial (P w \chi w)}{\partial z} &= - \left[ c_p P \, \frac{\partial \pi}{\partial z} + \rho g + (2\,\pmb{\Omega} \times \rho \pmb{v})_{\perp} \right] , 
        \label{eqn:reform_num_gov_eqns_c} \\
        \frac{\partial P}{\partial t} + \nabla_\Vert \cdot (P \pmb{u}) + \frac{\partial (Pw)}{\partial z} &= 0.
        \label{eqn:reform_num_gov_eqns_d}%
    \end{align}%
    \label{eqn:reform_num_gov_eqns}%
\end{subequations}%

\subsubsection{Auxiliary perturbation variables}
To facilitate the formulation of a semi-implicit discretisation with respect to gravity and sound propagation, \citet{bk2019} made use of the perturbation of the Exner pressure $\pi$ and the potential temperature variable $\chi$ via
\begin{equation}
    \pi(t, \pmb{x}, z) = \pi^\prime(t, \pmb{x}, z) + \bar{\pi}(z) \qquad \text{and} \qquad \chi(t, \pmb{x}, z) = \chi^\prime(t, \pmb{x}, z) + \bar{\chi}(z),
\end{equation}
where the former equation is reproduced from \eqref{eqn:pi_perturb} and with
\begin{equation}
    \frac{\mathrm{d} \bar{\pi}}{\mathrm{d} z} = - \frac{g}{c_p}\bar{\chi} \qquad \text{and} \qquad \bar{\pi}(0)=1
    \label{eqn:hydro_balance}
\end{equation}
describing the hydrostatically balanced background.

Since, in dimensionless form, $P=\pi^{\gamma-1}$ is a function of the Exner pressure alone, see \eqref{eqn:P_eos}, and as $\bar{\pi}$ is time-independent, one finds from \eqref{eqn:reform_num_gov_eqns_d} that
\begin{equation}
    \left( \frac{\partial P}{\partial \pi} \right) \frac{\partial \pi^\prime}{\partial t} = - \nabla \cdot \left[ P(\pi) \pmb{v} \right],
    \label{eqn:P-pi_updt}
\end{equation}
whereas the perturbation potential temperature variable $\chi^\prime$ follows
\begin{equation}
    \frac{\partial (P\chi^\prime)}{\partial t} + \nabla_\Vert \cdot (P \pmb{u} \chi^\prime) + \frac{\partial (P w \chi^\prime)}{\partial z} = - P w \frac{\partial \bar{\chi}}{\partial z}.
    \label{eqn:chi_prime_evo_eqn}
\end{equation}
Discretisation of \eqref{eqn:P-pi_updt} and \eqref{eqn:chi_prime_evo_eqn} will facilitate the semi-implicit integration of the full variable equations \eqref{eqn:reform_num_gov_eqns}. As in \citet{bk2019}, \eqref{eqn:P-pi_updt} is solved redundantly parallel to \eqref{eqn:reform_num_gov_eqns_d}.

Following the notation introduced by \cite{smolarkiewicz2014}, we let
\begin{equation}
    \pmb{\Psi} = (\chi, \chi \pmb{u}, \chi w, \chi^\prime),
\end{equation}
and summarise \eqref{eqn:reform_num_gov_eqns} and \eqref{eqn:chi_prime_evo_eqn} as
\begin{subequations}
    \begin{align}
        \frac{\partial (P \pmb{\Psi})}{\partial t} + \mathscr{A}(\Psi; P\pmb{v}) &= Q(\pmb{\Psi};P),
        \label{eqn:compact_a} \\
        \frac{\partial P}{\partial t} + \nabla \cdot (P\pmb{v}) & = 0,
        \label{eqn:compact_b}
    \end{align}
    \label{eqn:compact}%
\end{subequations}%
while \eqref{eqn:P-pi_updt}, being identical to \eqref{eqn:compact_b}, is not listed separately. In \eqref{eqn:compact}, $\mathscr{A}$ represents the update from a suitable advection scheme, and $Q(\pmb{\Psi};P)$ describes the effect of the right-hand side of \eqref{eqn:reform_num_gov_eqns} on $\pmb{\Psi}$ given P.

\subsection{Semi-implicit time discretisation}
\subsubsection{Implicit midpoint pressure update and advective fluxes}
Equation \eqref{eqn:reform_num_gov_eqns_d} is discretised with the midpoint rule, i.e.,
\begin{equation}
    (P^{n+1}-P^n) = - \Delta t \, \tilde{\nabla} \cdot (P\pmb{v})^{n+1/2}.
    \label{eqn:midpt_rule}
\end{equation}
Note that $\tilde{\nabla}\cdot$ is the discrete approximation of the divergence operator.

The update step \eqref{eqn:midpt_rule} requires the advective fluxes at the half-time level, $(P\pmb{v})^{n+1/2}$. This is computed via an explicit advection followed by a linearly implicit discretisation of the remaining and potentially stiff terms. The advection substep is given by
\begin{subequations}
    \begin{align}
        (P\pmb{\Psi})^\# &= \mathscr{A}_{\rm 1st}^{\Delta t / 2} (\pmb{\Psi}^n; (P\pmb{v})^n), 
        \label{eqn:hts_exp_updt_a} \\
        P^\# &= P^n - \frac{\Delta t}{2} \tilde{\nabla} \cdot (P\pmb{v})^n,
        \label{eqn:hts_exp_updt_b}%
    \end{align}%
    \label{eqn:hts_updt}%
\end{subequations}%
where $\mathscr{A}_{\rm 1st}^{\Delta t / 2}$ corresponds to an at least first-order accurate advection scheme. 

A more detailed explanation on this half-time update is provided in \S\,3b1 of \citet{bk2019}.

The $(P\pmb{v})^{n+1/2}$ field is then obtained from an implicit Euler substep for the stiff subset of equations,
\begin{subequations}
    \begin{align}
        (P\pmb{\Psi})^{n+1/2} &= (P\pmb{\Psi})^\# + \frac{\Delta t}{2} Q(\psi^{n+1/2}; P^{n+1/2}), 
        \label{eqn:hts_imp_updt_a} \\
        P^{n+1/2} &= P^n - \frac{\Delta t}{2} \tilde{\nabla} \cdot (P\pmb{v})^{n+1/2}.
        \label{eqn:hts_imp_updt_b}
    \end{align}
    \label{eqn:hts_imp_updt}%
\end{subequations}%
Here, \eqref{eqn:hts_imp_updt_b} is an implicit Euler update and thus is part of the implicit midpoint rule for $P$. The relation between $P$ and $\pi$ is approximated through a linearisation of the equation of state \eqref{eqn:P_eos},
\begin{equation}
    P^{n+1/2} = P^n + \left( \frac{\tilde{\partial} P}{\tilde{\partial} \pi} \right)^\# \left( \pi^{n+1/2} - \pi^n \right),
\end{equation}
which leads to a linear elliptic equation for $\pi^{n+1/2}$. Here, $\tilde{\partial}$ is the discrete partial derivative.

\subsubsection{Implicit trapezoidal rule for the advected quantities}
Given $(P\pmb{v})^{n+1/2}$, the semi-implicit time step for $\pmb{\Psi}$ reads
\begin{subequations}
\begin{align}
    (P\pmb{\Psi})^\ast &= (P\pmb{\Psi})^n + \frac{\Delta t}{2} Q(\pmb{\Psi}^n;P^n),
    \label{eqn:fts_updt_a} \\
    (P\pmb{\Psi})^{\ast\ast} &= \mathscr{A}_{\rm 2nd}^{\Delta t}(\pmb{\Psi}; (P\pmb{v})^{n+1/2}),
    \label{eqn:fts_updt_b} \\
    (P\pmb{\Psi})^{n+1} &= (P\pmb{\Psi})^{\ast\ast} + \frac{\Delta t}{2} Q(\pmb{\Psi}^{n+1};P^{n+1}),
    \label{eqn:fts_updt_c} \\
    P^{n+1} &= P^n - \Delta t \tilde{\nabla} \cdot (P\pmb{v})^{n+1/2}.
    \label{eqn:fts_updt_d}
\end{align}%
\label{eqn:fts_updt}%
\end{subequations}%
More details on the full-time update step in \eqref{eqn:fts_updt} is given in \S\,3b2 of \cite{bk2019}, while discretisation details of the second-order advection scheme in \eqref{eqn:fts_updt_b} are given \S\,4a and \S\,4b. Details on the discretisation of the semi-implicit integration of the stiff terms in \eqref{eqn:fts_updt_c} and \eqref{eqn:fts_updt_d}, and in particular the extension to support the non-traditional setting, are documented in Appendix~\ref{apx:extension}.

\section{Extension of the numerical model to support the non-traditional setting}
\label{apx:extension}
\subsection{Semi-implicit integration of the stiff terms}
The stiff terms in \eqref{eqn:reform_num_gov_eqns} are integrated by the implicit trapezoidal rule in time. Thus an explicit Euler step at the start of the time step is followed, after an advection corresponding to the left-hand side of \eqref{eqn:reform_num_gov_eqns}, by an implicit Euler step to complete a full-time update step. The implicit Euler scheme is also invoked in predicting the advective half-time level fluxes $(P\pmb{v})^{n+1/2}$ in \eqref{eqn:hts_imp_updt}. This crucial implicit Euler step, and an extension to support the non-traditional setting, is described in this Appendix.

As \eqref{eqn:reform_num_gov_eqns_a} and \eqref{eqn:reform_num_gov_eqns_d} are free of stiff terms, $\rho$ and $P$ are unchanged during the present implicit step. As a consequence, we have
\begin{subequations}
\begin{align}
    \frac{\partial U}{\partial t} &= - c_p (P\theta)^\circ \frac{\partial \pi^\prime}{\partial x} - \left( 2\, \Omega_y W - 2\, \Omega_z V\right),
    \label{eqn:imp_updt_a} \\
    \frac{\partial V}{\partial t} &= - c_p (P\theta)^\circ \frac{\partial \pi^\prime}{\partial y} - \left( 2\, \Omega_z U - 2\, \Omega_x W\right),
    \label{eqn:imp_updt_b} \\
    \frac{\partial W}{\partial t} &= - c_p (P\theta)^\circ \frac{\partial \pi^\prime}{\partial z} - g \frac{\tilde{\chi}}{\chi^\circ} - \left( 2\, \Omega_x V - 2\, \Omega_y U\right),
    \label{eqn:imp_updt_c} \\
    \frac{\partial \tilde{\chi}}{\partial t} &= -W \frac{\mathrm{d} \bar{\chi}}{\mathrm{d} z},
    \label{eqn:imp_updt_d} \\
    \left( \frac{\partial P}{\partial \pi} \right)^\circ \frac{\partial \pi^\prime}{\partial t} &= -\frac{\partial U}{\partial x} - \frac{\partial V}{\partial y} - \frac{\partial W}{\partial z},
    \label{eqn:imp_updt_e}%
\end{align}%
\label{eqn:imp_updt}%
\end{subequations}%
where $(U,V,W,\tilde{\chi}) = (Pu, Pv, Pw, P\chi^\prime)$, and $(P\theta)^\circ$, $\chi^\circ$, and $(\partial P / \partial \pi)^\circ$ are the data available when the implicit Euler solver is invoked. 
In the temporal semi-discretisation, this solver integrates
\begin{subequations}
\begin{align}
    U^{n+1} &= U^n - \Delta t \left[ c_p (P\theta)^\circ \frac{\tilde{\partial} \pi^{\prime\, n+1}}{\tilde{\partial} x} + \left( 2\, \Omega_y W^{n+1} - 2\,\Omega_z V^{n+1} \right) \right],
    \label{eqn:disc_imp_updt_a} \\
    V^{n+1} &= V^n - \Delta t \left[ c_p (P\theta)^\circ \frac{\tilde{\partial} \pi^{\prime\, n+1}}{\tilde{\partial} y} + \left( 2\, \Omega_z U^{n+1} - 2\,\Omega_x W^{n+1} \right) \right],
    \label{eqn:disc_imp_updt_b} \\
    W^{n+1} &= W^n - \Delta t \left[ c_p (P\theta)^\circ \frac{\tilde{\partial} \pi^{\prime\, n+1}}{\tilde{\partial} z} + g \frac{\tilde{\chi}^{n+1}}{\chi^\circ} + \left( 2\, \Omega_x V^{n+1} - 2\,\Omega_y U^{n+1} \right) \right],
    \label{eqn:disc_imp_updt_c} \\
    \tilde{\chi}^{n+1} &= \tilde{\chi}^{n} - \Delta t \, \frac{\tilde{\mathrm{d}} \bar{\chi}}{\tilde{\mathrm{d}} z} W^{n+1},
    \label{eqn:disc_imp_updt_d} \\
    \left( \frac{\tilde{\partial} P}{\tilde{\partial} \pi} \right)^\circ \pi^{\prime \, n+1} &= \left( \frac{\tilde{\partial} P}{\tilde{\partial} \pi} \right)^\circ \pi^{\prime \, n} - \Delta t \left( \frac{\tilde{\partial} U^{n+1}}{\tilde{\partial} x} - \frac{\tilde{\partial} V^{n+1}}{\tilde{\partial} y} - \frac{\tilde{\partial} W^{n+1}}{\tilde{\partial} z} \right).
    \label{eqn:disc_imp_updt_e}%
\end{align}%
\label{eqn:disc_imp_updt}%
\end{subequations}%
We note that the explicit Euler step in \eqref{eqn:fts_updt_a} entails solving \eqref{eqn:disc_imp_updt} with the time level $n+1$ replaced by the time level $n$ everywhere on the right-hand side of \eqref{eqn:disc_imp_updt}.

Now we let
\begin{equation}
    \nu = \Delta t \, N, \quad \omega_m = 2 \, \Delta t \, \Omega_m, \quad m \in \left\lbrace x,y,z\right\rbrace,
\end{equation}
with the local buoyancy frequency 
\begin{equation}
    N = \sqrt{-g \frac{1}{\chi^\circ} \frac{\mathrm{d} \bar{\chi}}{\mathrm{d}z}}
\end{equation}
in \eqref{eqn:bv_freq} now written in terms of the inverse potential temperature $\chi$. Furthermore, we let
\begin{equation}
    \pmb{U} = \begin{pmatrix}
        U \\ V \\ W
    \end{pmatrix},
    \quad
    \pmb{F} = \begin{pmatrix}
        0           &   -\omega_z   &   \omega_y    \\
        \omega_z    &   0           &   -\omega_x   \\
        -\omega_y   &   \omega_x    &   \nu^2
    \end{pmatrix},
    \quad
    \text{and}
    \quad
    \pmb{H} = \pmb{I} + \pmb{F},
\end{equation}
where $\pmb{I}$ is the identity matrix. Then the equation for the advective fluxes $(U,V,W)^{n+1}$ in \eqref{eqn:disc_imp_updt}, after the insertion of \eqref{eqn:disc_imp_updt_d} in \eqref{eqn:disc_imp_updt_c}, can be recast as,
\begin{equation}
    \pmb{U}^{n+1} = \pmb{U}^{n\ast} - \pmb{F} \, \pmb{U}^{n+1} - \Delta t \, c_p (P \theta)^{\circ} \tilde{\nabla} \pi^{\prime \, n+1},
\end{equation}
or, equivalently, 
\begin{equation}
    \pmb{U}^{n+1} = \pmb{H}^{-1} \left(\pmb{U}^{n\ast} - \Delta t \, c_p(P\theta)^{\circ} \tilde{\nabla} \pi^{\prime \, n+1 }\right),
    \label{eqn:Unp1_matrix_updt}
\end{equation}
where
\begin{equation}
    \pmb{U}^{n\ast} = \left( U^n, V^n, W^n - \Delta t \, g \frac{\tilde{\chi}^n}{\chi^\circ} \right)^T.
\end{equation}
Since
\begin{equation}
    \pmb{H} = \begin{pmatrix}
        1           &   -\omega_z   &   \omega_y    \\
        \omega_z    &   1           &   -\omega_x   \\
        -\omega_y   &   \omega_x    &   1 + \nu^2
    \end{pmatrix}
\end{equation}
and
\begin{equation}
    \det(\pmb{H}) = \omega_x^2 + \omega_y^2 + (1 + \nu^2)(1 + \omega_z^2),
\end{equation}
the inversion of $\pmb{H}$ leads to
\begin{equation}
    \pmb{H}^{-1} = \frac{1}{\det(\pmb{H})} \, \begin{pmatrix}
        (1 + \nu^2) + \omega_x^2 &
        (1 + \nu^2) \omega_z + \omega_x \omega_y &
        -\omega_y + \omega_z \omega_x 
        \\
        -(1 + \nu^2) \omega_z + \omega_x \omega_y &
        (1 + \nu^2) + \omega_y^2 &
        \omega_x + \omega_y \omega_z
        \\
        \omega_y + \omega_z \omega_x &
        -\omega_x + \omega_y \omega_z &
        1 + \omega_z^2
        \end{pmatrix}.
\end{equation}
Insertion of \eqref{eqn:Unp1_matrix_updt} into the pressure equation \eqref{eqn:disc_imp_updt_e} leads to a closed Helmholtz-type equation for $\pi^{\prime\,n+1}$,
\begin{equation}
    \left( \frac{\tilde{\partial} P}{\tilde{\partial} \pi} \right)^\circ \pi^{\prime \, n+1} - \Delta t^2 \, \tilde{\nabla} \cdot \left[ \pmb{H}^{-1} c_p (P \theta)^\circ \tilde{\nabla} \pi^{\prime \, n+1} \right] = \left( \frac{\tilde{\partial} P}{\tilde{\partial} \pi} \right)^\circ \pi^{\prime \, n} - \Delta t \, \tilde{\nabla} \cdot \left( \pmb{H}^{-1} \pmb{U}^{n\ast} \right).
    \label{eqn:helmholtz_prob}
\end{equation}
After the solution of \eqref{eqn:helmholtz_prob}, backward re-insertion into \eqref{eqn:disc_imp_updt_a}--\eqref{eqn:disc_imp_updt_d} yields $(U,V,W,\tilde{\chi})^{n+1}$. Details on computing the spatial derivatives on the right-hand side of \eqref{eqn:disc_imp_updt} are deferred to Section~4c3 of \citet{bk2019}.

\subsection{Solving for the Exner pressure perturbation increment}
For implementation purposes, we introduce the option of explicitly using the decomposition
\begin{equation}
    \pi^{\prime \, n+1} = \pi^{\prime \, n} + \delta \pi,
    \label{eqn:decomposition}
\end{equation}
such that only the time increment of the Exner pressure perturbation is updated in the solution of the Helmholtz-type problem in \eqref{eqn:helmholtz_prob}. Solution of the time increment only may improve the accuracy of the iterative solver for small time increments.

With the decomposition in \eqref{eqn:decomposition}, \eqref{eqn:helmholtz_prob} may be written as
\begin{equation}
    \left( \frac{\tilde{\partial} P}{\tilde{\partial} \pi} \right)^\circ \delta \pi - \Delta t^2 \, \tilde{\nabla} \cdot \left[ \pmb{H}^{-1} c_p (P \theta)^\circ \tilde{\nabla} \delta \pi \right] = - \Delta t \, \tilde{\nabla} \cdot \left( \pmb{H}^{-1} \pmb{U}^{n\ast\ast} \right),
    \label{eqn:helmholtz_prob_increment}
\end{equation}
where
\begin{equation}
    \pmb{U}^{n\ast\ast} = \pmb{U}^{n\ast} - \Delta t \, c_p (P\theta)^\circ \tilde{\nabla} \pi^{\prime \, n}.
\end{equation}
Upon obtaining the solution of $\delta \pi$, re-insertion into \eqref{eqn:decomposition} recovers $\pi^{n+1}$.

\section{Extension of the numerical model towards a well-balanced stable Lamb wave solution}
\label{apx:lamb_bal}
The equatorial Lamb wave in an isothermal atmosphere investigated numerically in Section~\ref{sec:num_experiments} is a special solution that has all the characteristics of a horizontally travelling acoustic wave, except that it is realised under the influence of gravity and with the associated strong vertical stratification of pressure and density. One important property of the wave is the zero vertical velocity.

In the context of simulating such a Lamb wave under the non-traditional setting, the ability of the numerical solver to maintain zero vertical velocity under the traditional approximation is crucial. This is because any instability would violate this property, and if the numerical solution of the Lamb wave in the traditional setting already has spurious numerically-induced vertical velocities, then these may trigger or interact with the unstable eigenmode in the non-traditional setting in uncontrolled ways.

Therefore, we derive in this Appendix a careful numerical treatment based on physical arguments that ensures a well-balanced Lamb wave solution in the traditional approximation. With the innovations presented in this Appendix, the linear solution of the traditional Lamb wave maintains a vertical velocity field up to machine accuracy for 10\,000 time steps (i.e., the maximum absolute amplitude of the vertical velocity field after the entire duration of the test simulation is $\sim10^{-16}$).

\subsection{Requirements for a well-balanced discretisation of the pressure equation}
Referring to \eqref{eqn:helmholtz_prob_increment} in Appendix~\ref{apx:extension}, let us consider the Lamb wave in the linear setting at the equator and under the traditional approximation for the Coriolis term. In this case,
\begin{equation}
    \pmb{H} \equiv \pmb{I}, 
    \qquad
    \frac{\tilde{\partial} P}{\tilde{\partial} \pi} \equiv \overline{\frac{\tilde{\partial} P}{\tilde{\partial} \pi}}, 
    \qquad
    (P\theta)^\circ \equiv \overline{P\theta},
    \qquad
    \text{and}
    \qquad
    R = -\Delta t \tilde{\nabla} \cdot (\bar{P} \pmb{u}^{n \ast \ast}),
\end{equation}
where the overline denotes the horizontally homogeneous background stratification.

From the analytical Lamb wave solution, we know that the Exner pressure perturbation $\pi^\prime$ is vertically homogenous, so that the entire Lamb wave structure satisfies the hydrostatic balance exactly. Therefore, we postulate a vertically homogeneous solution for $\delta \pi$ in \eqref{eqn:helmholtz_prob_increment} and then require the vertical dependencies of all terms in the equation to be exactly the same. This imposes a constraint on the discretisation that would guarantee the discrete solution for the time update of $\delta \pi$ to be vertically homogeneous up to machine accuracy. 

Taking into account that in the Lamb wave solution, we have
\begin{equation}
    \pmb{u}(t,x,z) = \bar{\theta}(z) \, \hat{\pmb{u}}(x-Ct),
    \label{eqn:horizontal_u}
\end{equation}
we must require that the coefficients in the operator on the left of \eqref{eqn:helmholtz_prob_increment} satisfy
\begin{equation}
    \overline{\frac{\tilde{\partial} P}{\tilde{\partial} \pi}} \sim \overline{P\theta}
    \label{eqn:op_constraint}
\end{equation}
exactly at the discrete level (with a suitable constant of proportionality), and that \eqref{eqn:horizontal_u} is satisfied accordingly.

That \eqref{eqn:op_constraint} is indeed satisfied by the Lamb wave solution at the continuum level is verified as follows: since $P$ is a function of $\pi$ alone, and since for an isothermal state with temperature $\pi\theta=T_0=\text{const.}$, we have
\begin{equation}
    \overline{\frac{dP}{d\pi}} = \frac{1}{\gamma - 1} \frac{\overline{P}}{\pi} = \frac{1}{\gamma -1} \frac{1}{T_0} \overline{P\theta}.
    \label{eqn:P_pi_balance}
\end{equation}
Moreover, we have from \eqref{eqn:horizontal_u}
\begin{equation}
    \overline{P}\pmb{u} = \overline{P\theta} \hat{\pmb{u}}.
\end{equation}
Thus we can proceed to ensure that these relations also hold at the discrete level. This is not a triviality because on the one hand, $P$ and $\pi$ are just functions of the pressure, but on the other hand, $\theta$ not only encodes the constant temperature but also determines the hydrostatic balance of the background state via \eqref{eqn:hydro_balance}. As a consequence, we have to invent a well-balanced discretisation of the hydrostatic relationship to guarantee that the distribution of $\bar{\theta}$ is synchronised with the distribution of $\bar{\pi}$ and $\overline{P}$.

\subsection{Well-balanced hydrostatic background state}
The starting point for defining the discrete background state of the isothermal atmosphere is the analytical solution for the Exner pressure, which is derived from
\begin{equation}
    c_p\theta \frac{d\bar{\pi}}{dz} = c_p T_0 \frac{d \ln{\bar{\pi}}}{dz} = -g, \qquad \bar{\pi}(0) = \pi_0 = 1.
\end{equation}
Here, we again use $\pi\theta=T$. The exact solution is
\begin{equation}
    \bar{\pi}(z) = \pi_0 \exp\left( - \frac{z}{H_\pi} \right), \quad \text{where} \quad H_{\pi} = \frac{c_p T_0}{g}.
\end{equation}
The first step in setting up the discrete background is to impose these exact relationships at the nodes of the computational grid, that is,
\begin{equation}
    \bar{\pi}_{j+1/2} = \bar{\pi}(z_{j+1/2}) \quad \text{and} \quad \bar{\theta}_{j+1/2} = \frac{T_0}{\bar{\pi}_{j+1/2}}.
    \label{eqn:pi_half}
\end{equation}

In the numerical scheme used here, the pressure gradients are approximated as 
\begin{equation}
    (\tilde{\nabla}\pi)_{i,j} = \frac{1}{|\mathcal{C}_{i,j}|} \mathop{\widetilde\oint_{\partial \mathcal{C}_{i,j}}} \pi \pmb{n} \, d\sigma,
\end{equation}
where $C$ is the control volume and the boundary integral is evaluated using the trapezoidal rule along the edges of the control volumes. This translates into the following discretisation of the hydrostatic balance,
\begin{equation}
    \frac{1}{\Delta z} \left( \bar{\pi}_{j+1/2} - \bar{\pi}_{j-1/2} \right) = - \frac{g}{c_p \bar{\theta}_j} \quad \text{i.e., we set} \quad \bar{\theta}_j = -\frac{g \Delta z}{c_p (\bar{\pi}_{j+1/2} - \bar{\pi}_{j-1/2})},
\end{equation}
with $\bar{\pi}_{j+1/2}$ from \eqref{eqn:pi_half}.
Furthermore, we set for the cell-centred pressure variables,
\begin{equation}
    \bar{\pi}_j = \frac{T_0}{\bar{\theta}_j}, \quad \text{and} \quad \overline{P}_j = (\bar{\pi}_j)^{\frac{1}{\gamma-1}}.
\end{equation}

The consequence is that, now,
\begin{equation}
    \left(\frac{\overline{dP}}{d\pi}\right)_j := \frac{1}{\gamma-1} \left( \overline{\frac{P}{\pi}} \right)_j = \frac{1}{\gamma-1} \frac{1}{T_0} \overline{P} \, \bar{\theta}, 
\end{equation}
which matches the requirement from \eqref{eqn:P_pi_balance}. All the other thermodynamic variables stored in the background states at the nodes and cell centres are computed from $(\bar{\pi}_{j+1/2}, \bar{\theta}_{j+1/2})$ and $(\bar{\pi}_{j}, \bar{\theta}_{j})$ as given above.

\subsection{Well-balanced extrapolation to the ghost cells}
Another point to be considered is the discretisation of the pressure equation \eqref{eqn:helmholtz_prob_increment} on the nodes that reside on the top and bottom boundaries. Taking into account only that half of the related dual cells overlaps with the flow domain, the original numerical scheme reduces to first order here. This gives rise to sizeable perturbations on the order of the truncation error of the scheme which includes unwanted vertical velocities. Applying the full stencil at these nodes by accessing the first outer row of ghost cells in formulating the flow divergence and Poisson stencil will guarantee second order accuracy and is, therefore, preferable. Yet, this is not sufficient to suppress vertical velocities down to machine accuracy, unless the extrapolation of flow states to the ghost cells is done carefully so as to maintain exact balance. This is the topic of the present subsubsection.

In fact, the states in the ghost cells must be obtained by an extrapolation from inside the domain that is analytically compatible with the Lamb wave structure. The relevant facts are as follows.
\begin{subequations}\label{eq:LambFacts}
\begin{enumerate}
\item Deviations of the Exner pressure from its background are vertically homogeneous, i.e., 
\begin{equation}\label{eq:LambFactsPi}
\frac{\partial (\pi - \bar{\pi})}{\partial z} = 0,
\end{equation}
\item the potential temperature equals its background stratification at all times, 
\begin{equation}\label{eq:LambFactsTheta}
\theta(t,x,z) = \bar{\theta}(z),
\end{equation}
\item the horizontal velocity divided by the potential temperature is vertically homogeneous, 
\begin{equation}\label{eq:LambFactsU}
\frac{\partial }{\partial z} \left(\frac{\pmb{u}}{\theta}\right) = 0,
\end{equation}
\item and the vertical velocity is zero,
\begin{equation}\label{eq:LambFactsW}
w(t,x,z) = 0.
\end{equation}
\end{enumerate}
We use these facts in constructing ghost cell values of the solution when needed. 
\end{subequations}

With $J$ denoting the vertical index of the top layer of control volumes within the flow domain and $i$ the horizontal index of a cell, we let
\begin{subequations}\label{eq:LambVerticalBCs}
\begin{align}
\label{eq:LambVerticalBCsPi}
\pi_{i,J+1} 
  & = \bar{\pi}_{J+1} + (\pi_{i,J} - \bar{\pi}_J),
    \\
\label{eq:LambVerticalBCsTheta}
\theta_{i,J+1}
  & = \bar{\theta}_{J+1},
    \\
\label{eq:LambVerticalBCsU}
\pmb{u}_{i,J+1} 
  & = \frac{\bar{\theta}_{J+1}}{\bar{\theta}_{J}} \pmb{u}_{i,J},
    \\
\label{eq:LambVerticalBCsW}
w_{i,J+1} 
  & = -\bar{\rho}_{i,J}\frac{\bar{\theta}_{i,J}}{\bar{\theta}_{i,J+1}} w_{i,J}.
\end{align}
The equations \eqref{eq:LambVerticalBCsPi}--\eqref{eq:LambVerticalBCsU} may be analogously applied to the bottom boundary, while 
\begin{equation}
w_{i,I-1} 
  = -\bar{\rho}_{i,I-1}\frac{\bar{\theta}_{i,I}}{\bar{\theta}_{i,I-1}} w_{i,I},
  \label{eq:LambVerticalBCsW_bot}
\end{equation}%
\label{eqn:bal_BCs}%
\end{subequations}%
where $I$ denotes the vertical index of the bottom layer of control volumes within the flow domain. All other variables are computed from this complete set. 

Eqs. \eqref{eq:LambVerticalBCsPi}--\eqref{eq:LambVerticalBCsU} should be clear from \eqref{eq:LambFactsPi}--\eqref{eq:LambFactsU}, while the handling of the vertical velocity at the top boundary in \eqref{eq:LambVerticalBCsW} ensures that the $P$-flux across the top boundary is zero. For the handling of the vertical velocity across the bottom boundary in \eqref{eq:LambVerticalBCsW_bot}, a re-weighting by the background density yields the most stable Lamb wave run.

\section{Stability of the Lamb wave solution under the traditional approximation}
\label{apx:energy}
This subsection aims to demonstrate that the numerical model introduced in Appendix~\ref{apx:num_model} and extended in Appendices~\ref{apx:extension} and \ref{apx:lamb_bal} can handle the stable Lamb wave configurations over longer-time simulations. Therefore, the onset of the instabilities in the numerical experiments of Section~\ref{sec:num_experiments} may be attributed solely to the experimental setup motivated by the theoretical developments in Section~\ref{sec:theory}.

In this subsection, two stable configurations of the Lamb wave are investigated. The initial condition for both runs is the Lamb wave solution, and rigid wall boundaries are applied to the top and the bottom. However, a non-traditional Coriolis strength of $\Omega_y=7.292\times10^{-5}$\,s$^{-1}$ is applied to one of the two runs for the duration of the numerical time integration after the initialisation of the fields at $t=0$\,s. This run is denoted by the label (LW-NT) and serves to demonstrate that even in the non-traditional setting, energy is nevertheless largely conserved if the experimental setup allows for it. The fully traditional stable Lamb wave run is denoted by (LW).
\begin{figure}[!h]
    \centerline{\includegraphics[width=1.0\textwidth]{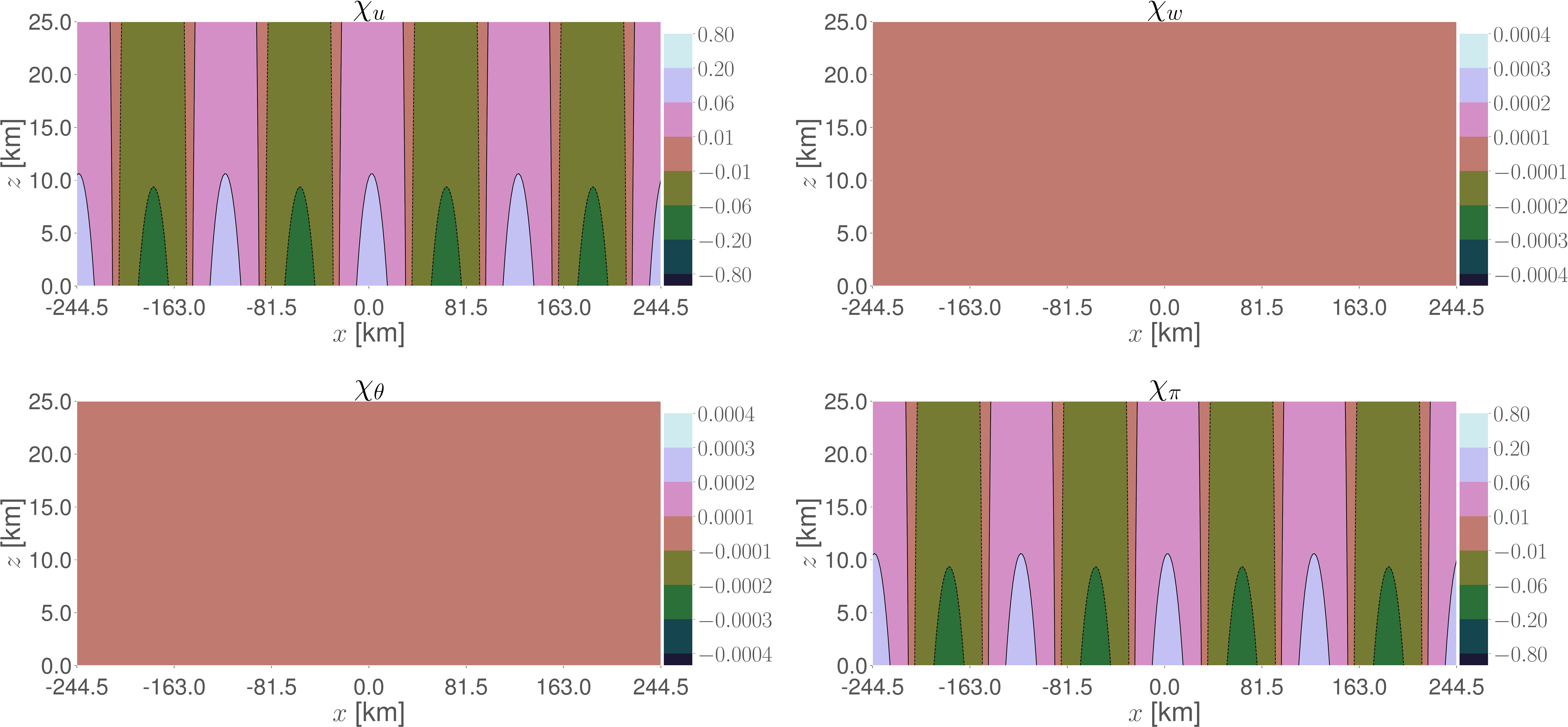}}
    \caption{Output of the fields for the transformed prognostic variables at $t=10.0$\,hrs for the configuration without the effect of the Coriolis force, i.e., $\Omega_y = 0.0$ corresponding to the LW traditional approximation run. Contour units are as in Figure~\ref{fig:trad_ICs}.}
    \label{fig:tradFinal}
\end{figure}
Figure~\ref{fig:tradFinal} shows the fields of the transformed prognostic variables of the perturbation for the LW case. The fields depicted are at $36\,000$\,s\,$= 10$\,hrs which is a significantly longer time scale in comparison with the time scale of the unstable mode as measured, for instance, by the doubling rate of 15\,min. 
The fields look almost identical to the initial condition in Figure~\ref{fig:trad_ICs}.
This is to be expected as the leading-order eigenvector is nothing but a Lamb wave in the traditional approximation which propagates indefinitely without change in the inviscid atmosphere.
\begin{figure}[!h]
    \centerline{\includegraphics[width=1.0\textwidth]{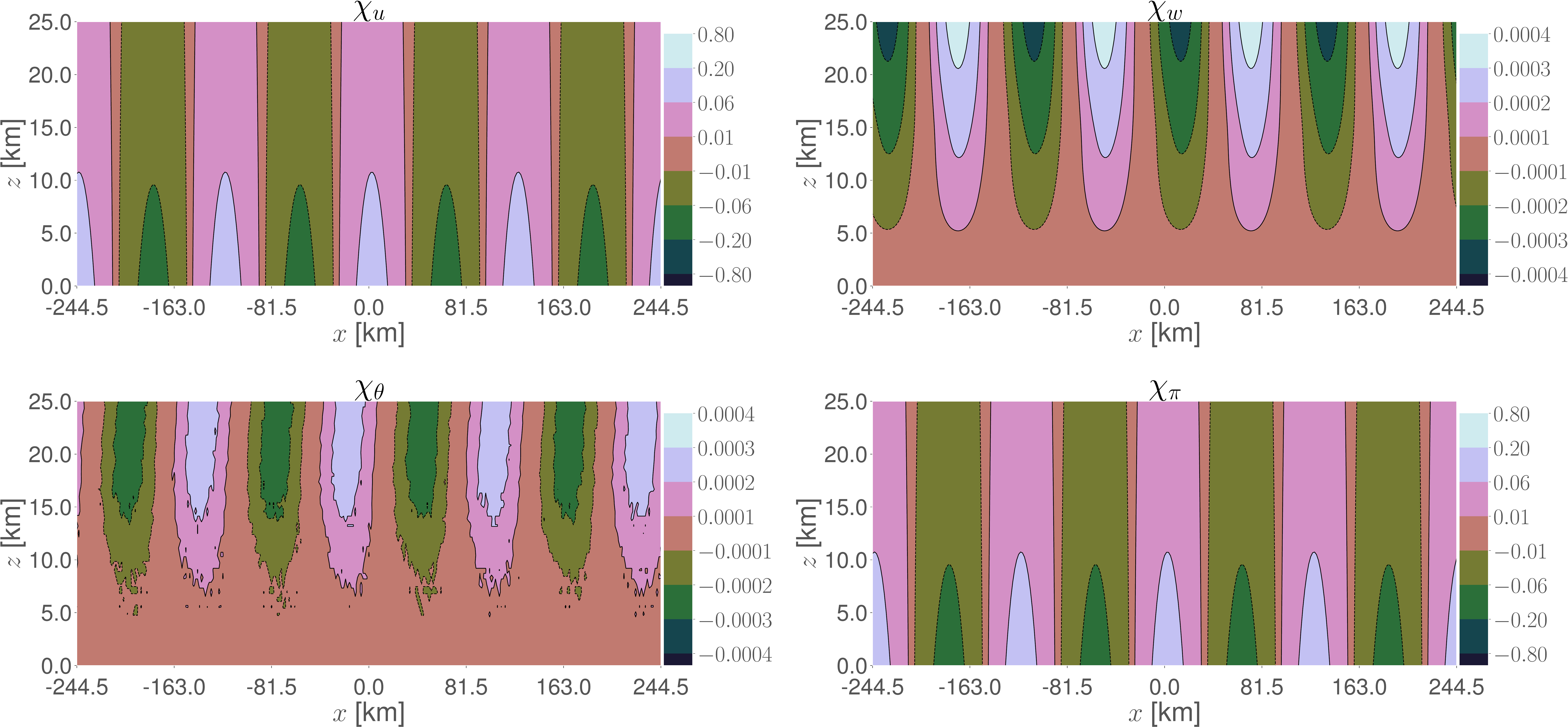}}
    \caption{Output of the fields for the transformed prognostic variables at $t=10.0$\,hrs for the LW-NT configuration. Contour units are as in Figure~\ref{fig:trad_ICs}.}
    \label{fig:hybridFinal}
\end{figure}
For the LW-NT run shown in Figure~\ref{fig:hybridFinal}, the $\chi_u$ and $\chi_\pi$ runs are similar to those of the fully traditional LW run. However, additional dynamics evolve in the $\chi_w$ and $\chi_\theta$ fields, and these fields are no longer approximately zero. Note that the colour scales in the $\chi_w$ and $\chi_\theta$ panels of Figures~\ref{fig:tradFinal} and \ref{fig:hybridFinal} are two orders of magnitude smaller than those presented in Figures~\ref{fig:trad_ICs} and~\ref{fig:unstable_ICs} of Section~\ref{sec:num_experiments}. This is to highlight that the amplitudes in the fields $\chi_w$ and $\chi_\theta$ remain small in both runs investigated here while also highlighting the effects of the non-traditional Coriolis on the flow dynamics in the LW-NT run.
\begin{figure}[!h]
    \centerline{\includegraphics[width=0.55\textwidth]{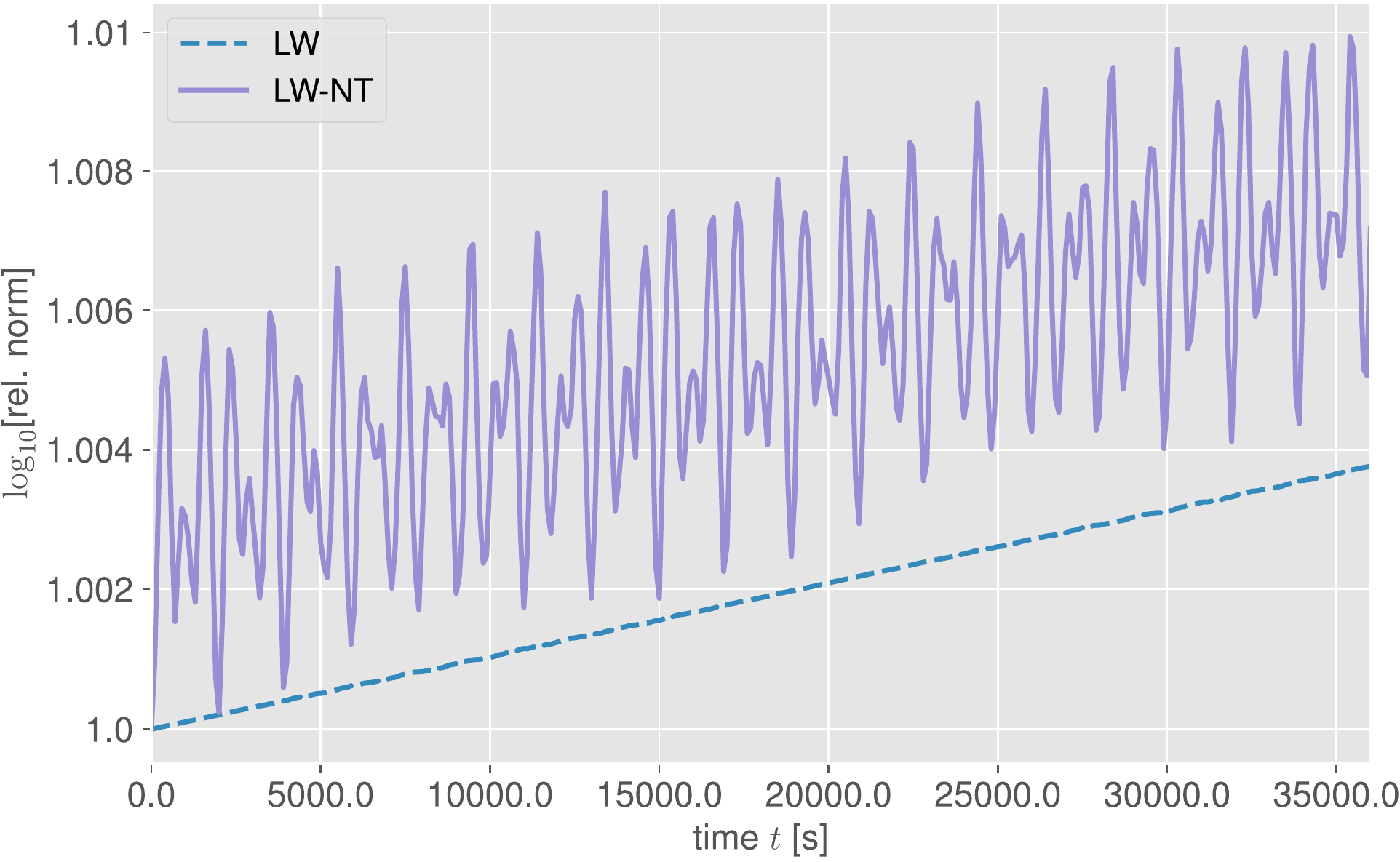}}
    \caption{Semi-logarithmic plot of the relative norm against time. The LW run with the traditional approximation (blue dashed curve), and the LW-NT run in the non-traditional setting (purple curve).}
    \label{fig:stabGrowth}
\end{figure}
How well the numerical model conserves the energy of the stable Lamb wave solutions may be seen in Figure~\ref{fig:stabGrowth}. Here, the relative norm, quantifying the relative energy, is depicted against time for the traditional LW (blue dashes) and non-traditional LW-NT (purple) runs. The traditional Lamb wave exhibits a slight increase in the energy at $\sim0.4\%$ after 10\,hrs, or $\sim0.04\%$ after an hour. This slight increase in the energy is negligible in comparison to the 1~hour outputs of the non-traditional Lamb-wave-like runs investigated in Section~\ref{sec:num_experiments}, which has increased by two orders of magnitude within this time period. 

By artificially including the non-traditional Coriolis in the LW-NT run, the relative energy of the system begins to fluctuate, as seen in the purple curve. However, notwithstanding the initial jump in the relative energy, the energy of the LW-NT Lamb wave is also largely conserved up to $\sim0.4\%$ after 10\,hrs. The numerical model therefore conserves the energy of a non-traditional Coriolis run if the boundary conditions were to allow for it.

\newpage
%%%%%%%%%%%%%%%%%%%%%%%%%%%%%%%%%%%%%%%%%%%%%%%%%%%%%%%%%%%%%%%%%%%%%
% REFERENCES
%%%%%%%%%%%%%%%%%%%%%%%%%%%%%%%%%%%%%%%%%%%%%%%%%%%%%%%%%%%%%%%%%%%%%
% Make your BibTeX bibliography by using these commands:
\bibliographystyle{abbrvnat}
\bibliography{libraryRay,libraryMark}

\end{document}

%% file: SketchOfSetup.tex
%% Creator: Inkscape 1.1.1 (3bf5ae0d25, 2021-09-20, custom), www.inkscape.org
%% Accompanies image file 'SketchOfSetup.pdf' (pdf, eps, ps)
\begingroup%
  \makeatletter%
  \providecommand\color[2][]{%
    \errmessage{(Inkscape) Color is used for the text in Inkscape, but the package 'color.sty' is not loaded}%
    \renewcommand\color[2][]{}%
  }%
  \providecommand\transparent[1]{%
    \errmessage{(Inkscape) Transparency is used (non-zero) for the text in Inkscape, but the package 'transparent.sty' is not loaded}%
    \renewcommand\transparent[1]{}%
  }%
  \providecommand\rotatebox[2]{#2}%
  \newcommand*\fsize{\dimexpr\f@size pt\relax}%
  \newcommand*\lineheight[1]{\fontsize{\fsize}{#1\fsize}\selectfont}%
  \ifx\svgwidth\undefined%
    \setlength{\unitlength}{209.76377953bp}%
    \ifx\svgscale\undefined%
      \relax%
    \else%
      \setlength{\unitlength}{\unitlength * \real{\svgscale}}%
    \fi%
  \else%
    \setlength{\unitlength}{\svgwidth}%
  \fi%
  \global\let\svgwidth\undefined%
  \global\let\svgscale\undefined%
  \makeatother%
  \begin{picture}(1,0.7027027)%
    \lineheight{1}%
    \setlength\tabcolsep{0pt}%
    \put(1.83607885,3.50326038){\color[rgb]{0,0,0}\makebox(0,0)[lt]{\begin{minipage}{0.5938652\unitlength}\raggedright  \end{minipage}}}%
    \put(0.70545853,0.2949068){\color[rgb]{0,0,0}\makebox(0,0)[lt]{\lineheight{0}\smash{\begin{tabular}[t]{l}$\Omega$\end{tabular}}}}%
    \put(0.26151868,0.02451709){\color[rgb]{0,0,0}\makebox(0,0)[lt]{\lineheight{0}\smash{\begin{tabular}[t]{l}$x$\end{tabular}}}}%
    \put(0.15255256,0.61838264){\color[rgb]{0,0,0}\makebox(0,0)[lt]{\lineheight{0}\smash{\begin{tabular}[t]{l}$z$\end{tabular}}}}%
    \put(0.17434576,0.30510498){\color[rgb]{0,0,0}\makebox(0,0)[lt]{\lineheight{0}\smash{\begin{tabular}[t]{l}$\pmb{v}$\end{tabular}}}}%
    \put(0.4603818,0.61565845){\color[rgb]{0,0,0}\makebox(0,0)[lt]{\lineheight{0}\smash{\begin{tabular}[t]{l}$2\pmb{\Omega}\times\pmb{v}$\end{tabular}}}}%
    \put(0,0){\includegraphics[width=\unitlength]{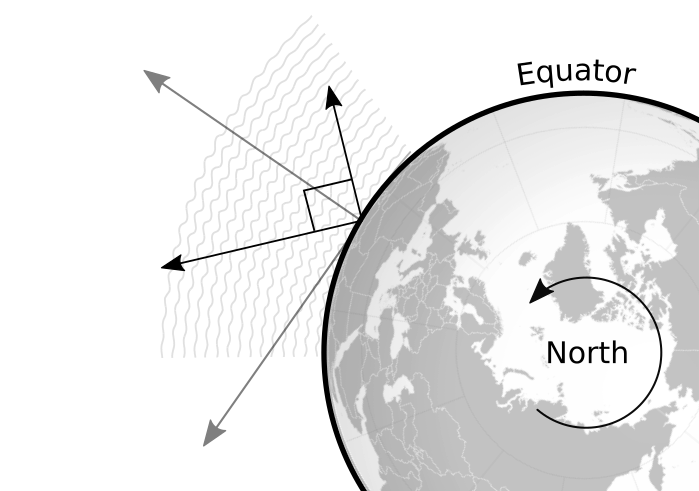}}%
  \end{picture}%
\endgroup%